\documentclass[11pt]{article}
\usepackage{latexsym,epsfig,euscript,amssymb,amsmath,verbatim,color,subfigure}
\newcommand{\beq}{\begin{equation}}
\newcommand{\eeq}{\end{equation}}
\newcommand{\beqs}{\begin{eqnarray}}
\newcommand{\eeqs}{\end{eqnarray}}
\newcommand{\bit}{\begin{itemize}}
\newcommand{\eit}{\end{itemize}}
\newcommand{\bce}{\begin{center}}
\newcommand{\ece}{\end{center}}
\newcommand{\ben}{\begin{enumerate}}
\newcommand{\een}{\end{enumerate}}
\newcommand{\tr}{\mathrm{tr}}
\newcommand{\hc}{\mathrm{h.c.}}

\newcommand{\nn}{\nonumber}

\newcommand{\GHC}{G_{\mathrm{HC}}}
\newcommand{\GSM}{G_{\mathrm{SM}}}
\newcommand{\NHC}{N_{\mathrm{HC}}}
\newcommand{\GCUS}{G_{\mathrm{cus.}}}
\newcommand{\GF}{G_{\mathrm{F}}}
\newcommand{\HF}{H_{\mathrm{F}}}
\newcommand{\sh}{{\mathrm s}_h}
\newcommand{\ch}{{\mathrm c}_h}
\newcommand{\sw}{{\mathrm s}_{\mathrm w}}
\newcommand{\cw}{{\mathrm c}_{\mathrm w}}

\begin{document}

\pagestyle{empty}

\begin{center}

{\LARGE{\bf UV Completions of Partial Compositeness: \\ \bigskip
The Case for a $SU(4)$ Gauge Group}}

\vspace{1.8cm}

{\large{Gabriele Ferretti}}

\vspace{1.5cm}

{\it Department of Fundamental Physics, \\
Chalmers University of Technology, \\
Fysikg{\aa}rden 1, 41296 G\"oteborg, Sweden\\
{\tt ferretti@chalmers.se}}

\vspace{2.cm}

\begin{minipage}[h]{14.0cm}

\bce {\bf \large Abstract} \ece

\medskip
We present a model of partial compositeness arising as the IR limit of a $SU(4)$ gauge theory with only fermionic matter. This group is one of the most promising ones among a handful of possible choices allowing a symmetry breaking pattern incorporating custodial symmetry and a top partner candidate, while retaining asymptotic freedom. It is favored for not giving rise to lepto-quarks or Landau poles in the SM gauge couplings. The minimal UV theory consists of five hyperfermions in the anti-symmetric representation and three in the fundamental and anti-fundamental.
The IR theory is centered around the coset $SU(5)/SO(5)$, with top partners in the fundamental of $SO(5)$, giving rise to one composite fermion of electric charge $5/3$, three of charge $2/3$ and one of charge $-1/3$.
Electro-Weak symmetry breaking occurs via top-quark-driven vacuum misalignment. The top quark mass is generated via the mechanism of partial compositeness, while the remaining fermions acquire a mass via a standard quadratic coupling to the Higgs. We compute the top and bottom quark mass matrix and the Electro-Weak currents of the composite fermions. The model does not give rise to unacceptably large deviations from the SM $Z\to b \bar b$ decay width.
\end{minipage}
\end{center}
\newpage

%  Resetting of counters
\setcounter{page}{1} \pagestyle{plain} \renewcommand{\thefootnote}{\arabic{footnote}} \setcounter{footnote}{0}

\section{Introduction}

The discovery~\cite{LHC} of a 126 GeV Higgs boson~\cite{BEH}, together with our expectations from effective field theory, points to the existence of new states and enlarged symmetries at the LHC scale. While nowadays some degree of fine tuning seems almost unavoidable in any incarnation of this idea, due to the fierce direct and indirect experimental constraints, one possibility that still remains is the existence of a new strongly coupled gauge theory at a scale much below the GUT scale.

For this idea still to be viable today, some specific dynamical mechanisms must occur. Among the few possibilities, we concentrate on the following scenario, generally known under the name of ``partial compositeness'':
\bit
\item[i)] First, the Higgs boson arises as a pseudo Nambu-Goldstone boson (pNGB) of a broken global symmetry and condenses at the EW scale $v= 246$~GeV via a ``misalignment'' mechanism~\cite{Kaplan:1983fs}. This guarantees that the corrections to the $S$ parameter are suppressed by a factor $v^2/f^2 \ll 1$, with $f$ the decay constant of the pNGB.
\item[ii)] Second, the top quark, (and possibly other fermions), acquires a mass by mixing with a composite state of the same quantum numbers~\cite{Kaplan:1991dc}. This helps in suppressing flavor changing neutral currents (FCNC) and CP violating terms without reintroducing a large fine-tuning.
\eit
Many works on this subject start with a phenomenological lagrangian with the desired properties and use the CCWZ formalism~\cite{CCWZ} to describe the interactions. Attempts to derive this lagrangian from an underlying model have been mostly based on the idea of extra dimensions and holography.
We will not discuss these approaches in this paper and instead will refer to the many reviews~\cite{manyreviews} and references therein for the original literature. (We have been mostly following~\cite{Contino:2010rs}.)

Work on purely four-dimensional UV completions, based on some strongly coupled ``hypercolor'' (HC) group, has been hampered by the objective difficulty of constructing entirely satisfactory models giving rise to the two dynamical mechanisms above. One difficulty is in obtaining viable partners to all the Standard Model (SM) fermions. Another difficulty is in achieving realistic masses for those that do have a partner. One must require a mixing, schematically of the type $\bar q {\mathcal{O}}$, between a generic SM fermion $q$ and a composite state ${\mathcal{O}}$. In order for this mechanism to be effective, the scaling dimension of ${\mathcal{O}}$ must be close to $5/2$. This is easy to realize in the presence of elementary scalars $\phi$ in the HC theory as ${\mathcal{O}}\approx \phi\psi$ (where $\psi$ is a HC fermion), but the reappearance of scalars calls once again for an explanation. This strategy is being pursued in the context of supersymmetric theories in e.g.~\cite{Caracciolo:2012je}.

Purely fermionic UV completions require ${\mathcal{O}}$ to attain a large anomalous dimension. Apart for the exceptional case of an adjoint HC fermion $\psi$ that can combine with the HC field strength $F$ to give ${\mathcal{O}}\approx F_{\mu\nu}\gamma^{\mu\nu}\psi$ of perturbative dimension $7/2$, the other possibility, for generic irreps, is to have some HC invariant combination ${\mathcal{O}}\approx \psi_1\psi_2\psi_3$ of perturbative dimension $9/2$, requiring an anomalous dimension $\eta \approx - 2$. While this is a tall order, it is nevertheless more appealing than the corresponding requirement needed for the pNGB composite operator ${\mathcal{H}}$ in the case where SM fermion masses are obtained by a bilinear term $\bar q {\mathcal{H}} q$.
In this latter case~\cite{Luty:2004ye}, (see also~\cite{smallN, higher}\footnote{In particular, in~\cite{higher}, some higher dimensional irreps have been studied that will also appear in the present work.}),
the requirement on ${\mathcal{H}}$ is that it has scaling dimension close to $1$, but this is the free field limit for a boson and implies that the scaling dimension of ${\mathcal{H}}^\dagger {\mathcal{H}}$ cannot be much different from $2$, reintroducing the fine-tuning
problem~\cite{confo}. On the contrary, $5/2$ is safely above the free field case for a fermion and in any case it does not give rise to additional relevant perturbations. However, the idea~\cite{Luty:2004ye} may still be viable for the SM fermions other than the top quark and we will rely on this in our construction.

A purely fermionic model of this type was proposed in~\cite{Barnard:2013zea} based on a HC group $Sp(4)$ and some of its basic dynamical properties were studied. In~\cite{Ferretti:2013kya} we classified, purely on group theoretical grounds, the models that fulfill the requirements i) and ii) above, together with some extra simplifying conditions such as a simple HC group. In~\cite{Ferretti:2013kya} we made no attempt to study the dynamics of these models. In this work, we return to this issue and consider one of the most attractive models in the classification~\cite{Ferretti:2013kya}, based on a HC group $SU(4)$.

Given that, in the most favorable possible scenario, the LHC will find evidence for compositeness that can be fully described by the IR effective theory,
what is the interest in looking for UV completions now? One reason is that, in the strictly IR approach, one has no control over the possible group realizations, (both the coset and the irreps) of the theory and one is forced to guess or to scan over ``group theory space'' (see e.g.~\cite{DeSimone:2012fs, Carena:2014ria}). The UV completion can help pointing towards the most promising models. Equivalently, by considering what generic properties arise in the IR from a class of UV theories, one can test or rule out the whole UV class.

Let us summarize the organization and the main results of the paper.

In Section 2, we present the UV theory. We discuss its matter content, the pattern of symmetry breaking and the composition of the top partners in terms of the hyperfermions. We show that the theory does not give rise to leptoquarks or any scalar composite state in the triplet or sextet of color. We compute the modification to the SM $\beta$-functions and show that no Landau pole arises at low scales. There is an amusing coincidence where the SM couplings almost unify but the
scale at which it occurs is too small to be taken seriously and, at any rate, we know that new physics must arise before that to generate the needed couplings between the SM and the hypercolor sector.

Section 3 discusses the IR theory. We present the pNGB and top partner field content and argue that EW breaking proceeds as required. We then construct the relevant couplings between the SM and the composite fields. Due to the lack of potential partners for all SM fermions, partial compositeness is applied only to the top quark, and we propose that the remaining fields should be given a mass by standard quadratic interactions. We discuss what spurions should be used for this purpose. We construct the EW currents and the derivative couplings of the composite fermions. Here we find a happy circumstance when it comes to the $Z\to b \bar b$ decay. The irreps involved are such that the decay is safe from large corrections~\cite{Agashe:2006at} arising from the composite partners. We also show this explicitly by going to the $b$ mass eigenstates.

In Section 4, we conclude with a short discussion and briefly review the current experimental status.

The main omission in this work is that we do not attempt to show that the anomalous dimensions for the composite operators are sufficient to realize a realistic mass spectrum, although arguments in favor of this possibility have been recently proposed in~\cite{Barnard:2013zea} for a similar model. Convincing evidence on this issue can only come via lattice simulations or a detailed analysis of the OPE that is beyond the scope of this paper. We also do not speculate on what physics could give rise to the required four-fermi couplings at a much higher scale.

\section{The UV theory}

In~\cite{Ferretti:2013kya} we searched for gauge theories with fermionic matter allowing a spontaneous global symmetry breaking pattern $\GF/\HF$ compatible with custodial symmetry: $\HF \supset \GCUS \supset \GSM$. (Having defined $\GCUS = SU(3)_c \times SU(2)_L \times SU(2)_R \times U(1)_X$ and $\GSM = SU(3)_c\times SU(2)_L\times U(1)_Y$.) We further required the presence of one Higgs doublet $\GF/\HF \ni
({\mathbf{1}},{\mathbf{2}},{\mathbf{2}})_0$ of $\GCUS$ and a composite fermionic trilinear partner for at least the third generation $\GSM$ fermions $Q_L \in ({\mathbf{3}}, \mathbf{2})_{1/6} $ and $t_R \in (\mathbf{3}, \mathbf{1})_{2/3}$.

We restricted the search to asymptotically free theories with a simple HC group $\GHC$ and at most three inequivalent types of fermionic irreps.
One could enlarge the class of theories, but the restricted class above already captures all the desired features.
The solutions to the constraints above where presented in Tables~2 and~3 of~\cite{Ferretti:2013kya} and included the model presented in~\cite{Barnard:2013zea}. One can classify these models in various way. One possibility is to divide them according to the breaking of the global symmetry giving rise to the pNGB's. The two custodial cosets arising contain either $SU(n)/Sp(n)$ or $SU(n)/SO(n)$, with $n = 4$ and $5$ being the minimal choice respectively.

Another distinction that can be made between them is whether they allow for composite scalars in the $\mathbf 3$ or $\mathbf 6$ of $SU(3)_c$. According to ones expectations, these are either exciting predictions or potential problems for these models and their role has been discussed in e.g.~\cite{oai:arXiv.org:0910.1789}. The model~\cite{Barnard:2013zea} contains such states originating from fermionic bilinears in the theory. We chose to work with theories that \emph{do not} give rise to such states and this restricts the number of solutions considerably. In fact, this requirement, together with the requirement that the new hyperfermions do not give rise to Landau poles too close to the EW scale, essentially singles out one solution, presented here in Table~\ref{SU4content}, based on the hypercolor group $\GHC=SU(4)$ which will be the focus of this paper. It is interesting to notice that $SU(4)$ is the only unitary group allowing this construction.

\begin{table}
  \centering
  \begin{tabular}{c|c|c|c|c|c|c|}
    % after \\: \hline or \cline{col1-col2} \cline{col3-col4} ...
     \multicolumn{1}{c}{} & \multicolumn{1}{c}{$\GHC$}& \multicolumn{5}{c}{$\GF$}\\
     \multicolumn{1}{c}{} & \multicolumn{1}{c}{$\overbrace{\phantom{aaaaa}}$}& \multicolumn{5}{c}{$\overbrace{\phantom{aaaaaaaaaaaaaaaaaaaaaaaaaaaaaaaaaa}}$}\\
      & $SU(4)$ & $SU(5) $& $SU(3)$ & $SU(3)'$ & $U(1)_X$ & $U(1)'$  \\
      \hline
    $\psi$ & $\mathbf{6}$ & $\mathbf{5}$ & $\mathbf{1}$ & $\mathbf{1}$ & $0$ & $-1$ \\
    $\chi $ &$\mathbf{4}$ & $\mathbf{1}$ & $\mathbf{3}$ & $\mathbf{1}$ & $-1/3$ & $5/3$\\
    $\tilde\chi $ & $\bar{\mathbf{4}}$ & $\mathbf{1}$ & $\mathbf{1}$ & $\bar{\mathbf{3}}$ & $1/3$ & $5/3$ \\
    \hline
  \end{tabular}
  \caption{\small The fermions of the UV theory studied in this paper. They are to be thought of as two-component left-handed objects. Later, when discussing the low energy phenomenological lagrangian, we will find it more convenient to revert to four-component notation. $\GHC$ is the hypercolor gauge group and $\GF$ the global symmetry group before symmetry breaking.}\label{SU4content}
\end{table}

\subsection{Field content of the UV theory}
\begin{table}
  \centering
  \begin{tabular}{c|c|l|}
    % after \\: \hline or \cline{col1-col2} \cline{col3-col4} ...
       Object &$SU(2)_L\times SU(2)_R$ &$SU(5) \times SU(3)\times SU(3)' \times U(1)_X \times U(1)'$\\
      \hline
      $\chi\psi\chi$ & (1/2, 0) & \hspace{2.7cm} $(\mathbf{5}, \bar{\mathbf{3}}, \mathbf{1})_{(-2/3,~7/3)}$ \\
                     & (1/2, 0) & \hspace{2.7cm} $(\mathbf{5}, \mathbf{6}, \mathbf{1})_{(-2/3,~7/3)}$ \\
                     & (3/2, 0) & \hspace{2.7cm} $(\mathbf{5}, \mathbf{6}, \mathbf{1})_{(-2/3,~7/3)}$ \\
      \hline
      $\tilde\chi\psi\tilde\chi$ & (1/2, 0) & \hspace{2.7cm} $(\mathbf{5}, \mathbf{1}, \mathbf{3})_{(2/3,~7/3)}$ \\
                                 & (1/2, 0) & \hspace{2.7cm} $(\mathbf{5}, \mathbf{1}, \bar{\mathbf{6}})_{(2/3,~7/3)}$ \\
                                 & (3/2, 0) & \hspace{2.7cm} $(\mathbf{5}, \mathbf{1}, \bar{\mathbf{6}})_{(2/3,~7/3)}$ \\
      \hline
      $\bar{\tilde\chi}\bar\psi\chi$ & (1/2, 0) & \hspace{2.7cm} $(\bar{\mathbf{5}}, \mathbf{3}, \mathbf{3})_{(-2/3,~1)}$ \\
                                     & (1/2, 1) & \hspace{2.7cm} $(\bar{\mathbf{5}}, \mathbf{3}, \mathbf{3})_{(-2/3,~1)}$ \\
      \hline
      $\bar{\chi}\bar\psi\tilde\chi$  & (1/2, 0) & \hspace{2.7cm} $(\bar{\mathbf{5}}, \bar{\mathbf{3}},
                                                            \bar{\mathbf{3}})_{(2/3,~1)\phantom{-}}$ \\
                                      & (1/2, 1) & \hspace{2.7cm} $(\bar{\mathbf{5}}, \bar{\mathbf{3}},
                                                            \bar{\mathbf{3}})_{(2/3,~1)\phantom{-}}$ \\
      \hline
      $\bar\chi\psi\bar\chi$ & (1/2, 0) & \hspace{2.7cm} $(\mathbf{5}, \mathbf{3}, \mathbf{1})_{(2/3,~-13/3)}$ \\
                             & (1/2, 1) & \hspace{2.7cm} $(\mathbf{5}, \bar{\mathbf{6}}, \mathbf{1})_{(2/3,~-13/3)}$ \\
      \hline
      $\bar{\tilde\chi}\psi\bar{\tilde\chi}$& (1/2, 0) & \hspace{2.7cm} $(\mathbf{5}, \mathbf{1}, \bar{\mathbf{3}})_{(-2/3,~-13/3)}$ \\
                                            & (1/2, 1) & \hspace{2.7cm} $(\mathbf{5}, \mathbf{1}, \mathbf{6})_{(-2/3,~-13/3)}$ \\
    \hline
  \end{tabular}
  \caption{\small The composite fermionic operators classified according to their Lorentz and flavor quantum numbers. For each operator there is a corresponding conjugate one. After symmetry breaking, they combine into vector-like operators that create spin 1/2 or spin 3/2 resonances out of the vacuum.}\label{compositefermions}
\end{table}

Since the full matter content in Table~\ref{SU4content} is non-chiral ($\mathbf{6}$ is a real irrep of $SU(4)$), the theory is manifestly free of gauge anomalies $\GHC^3$. The group
$\GF = SU(5)\times SU(3)\times SU(3)'\times U(1)_X\times U(1)'$ describes the flavor group free of ABJ anomalies $\GF\GHC^2$. The QCD color gauge group $SU(3)_c$ is realized as the diagonal subgroup of $SU(3)\times SU(3)'$, in perfect analogy with the flavor symmetries for the light quarks. The subgroup of $\GF$ that does not give rise to 't~Hooft anomalies $\GF^3$, and thus can be weakly gauged when coupled to the SM, is $\HF = SO(5)\times SU(3)_c \times U(1)_X$, containing the custodial group $\GCUS$ defined above.

The reason why it is not possible to build mesons (of any spin) bilinear in the $\chi, \tilde\chi$ fields and transforming in the $\mathbf 3$ or $\mathbf 6$ of $SU(3)_c$ is that $\chi$ and $\tilde\chi$ transform under a complex irrep of $\GHC$. In the present case, with the field content of Table~\ref{SU4content}, after reducing the
$({\mathbf{3}}, \bar{\mathbf{3}})$ of $SU(3)\times SU(3)'$ to color $SU(3)_c$, one can only construct color singlet/octet scalars\footnote{We will sometimes drop all indices to avoid cluttering the formulas when the contractions are obvious. For instance, the vector octet is the traceless part of $\chi^\dagger_{m a}\sigma^\mu\chi^{m b}$. In the paper, $m, n\dots $, $I,J\dots $, $a, b \dots $ and $a', b' \dots $ are $SU(4)$, $SU(5)$, $SU(3)$ and $SU(3)'$ indices respectively and the contraction over the Weyl indices is understood.} of type
$\tilde\chi \chi$, $\tilde\chi^\dagger \chi^\dagger$, or color singlet/octet vectors $\chi^\dagger \chi$,
$\tilde\chi^\dagger \tilde\chi$.

In Table~\ref{compositefermions} we list all fermionic $\GHC$ invariant that can be constructed with three elementary fields, together with their spin and $\GF$ flavor quantum numbers (later to be broken to $\HF$). This list includes the top quark partners that will be of interest in the remaining sections.

\subsection{Symmetry breaking in the UV theory}
Now we would like to argue that the pattern of symmetry breaking to be expected in this model is $\GF \to \HF$, with $\GF$ and $\HF$ given above, while leaving the hypercolor gauge group $\GHC=SU(4)$ unbroken. Since $\psi$ is in a real representation of $\GHC$, all the fermionic objects in Table~\ref{compositefermions} can be made massive by giving a mass to the $\psi$ fields. This means that none would be available to cancel the 't~Hooft anomalies~\cite{'tHooft:1980xb} associated to the $\GF/\HF$ generators, which should then be broken~\cite{Preskill:1981sr}. This patter of symmetry breaking is also consistent with the arguments of~\cite{VW}.

A more dynamical argument is an adaptation of the Nambu--Jona-Lasinio method~\cite{NJL} as done in~\cite{Barnard:2013zea}. The $\GHC$-invariant scalar bilinears that can be constructed are $\epsilon_{mnpq}\psi^{Imn}\psi^{Jpq}$, ~$\tilde\chi_{ma'}\chi^{ma}$ and their complex conjugates.

Introducing two auxiliary fields $M^{IJ} \equiv M^{JI}$ and $N_{a'}^a$, the fourth-order effective lagrangian can be written as
\beqs
     {\mathcal{L}}&\supset& -\frac{3}{2k} M^{IJ} M^\dagger_{IJ} - \frac{1}{2} M^{IJ}\epsilon^{mnpq} \psi^\dagger_{Imn}\psi^\dagger_{Jpq}
     - \frac{1}{2} M^\dagger_{IJ}\epsilon_{mnpq} \psi^{Imn}\psi^{Jpq}\label{L1}\\
     && - \frac{9}{k'}N^{\dagger a'}_a N_{a'}^a  - N_{a'}^a\tilde\chi^{\dagger m a'}\chi^\dagger_{ma}
     - N^{\dagger a'}_a\tilde\chi_{m a'}\chi^{ma} \nonumber
\eeqs
which, eliminating $M$ and $N$, reduces to
\beq
     {\mathcal{L}}\supset \frac{k}{6} \epsilon_{mnpq} \psi^{Imn}\psi^{Jpq} \epsilon^{m'n'p'q'} \psi^\dagger_{Im'n'}\psi^\dagger_{Jp'q'}+
                    \frac{k'}{9}\tilde\chi_{m a'}\chi^{ma}\tilde\chi^{\dagger na'}\chi^\dagger_{na}.
\eeq
The fields $M$ and $N$ can be reduced to non-negative diagonal form by Takagi and singular-value decomposition respectively
\beqs
     && M^{IJ}=\sum_K  \mu_K \Omega^I_K \Omega^J_K, \quad \psi^K = \Omega_I^K \psi'^I \label{decomposistion}\\
     && N_{a'}^a = \sum_b \nu_b {\tilde\Xi}_{a'}^b \Xi_b^a,
           \quad \chi^a = \Xi_b^a \chi'^b~\hbox{ and }~\tilde\chi_{a'}= {\tilde\Xi}_{a'}^b \tilde\chi_{b}.\nonumber
\eeqs
In (\ref{decomposistion}), $\Omega$, $\Xi$ and $\tilde\Xi$ are orthogonal matrices and the sum is indicated explicitly only when the contraction is not manifestly group-covariant. Using (\ref{decomposistion}), (\ref{L1}) becomes
\beqs
     {\mathcal{L}}&\supset&  \sum_{I=1}^5 -\frac{3}{2k} \mu_I^2 - \frac{1}{2} \mu_I \epsilon^{mnpq} \psi'^\dagger_{Imn}\psi'^\dagger_{Ipq}
     - \frac{1}{2} \mu_I \epsilon_{mnpq} \psi'^{Imn}\psi'^{Ipq}\label{L2}\\
     && \sum_{a=1}^3  - \frac{9}{k'}\nu_a^2 - \nu_a \tilde\chi'^{\dagger m a}\chi'^\dagger_{ma}
     - \nu_a \tilde\chi'_{m a}\chi'^{ma}. \nonumber
\eeqs
Integrating out the fermions, with $\Lambda$ interpreted as the $\GHC$ strong scale, gives
\beq
    {\mathcal{L}}\supset - \sum_{I=1}^5 V(\mu_I) - \sum_{a=1}^3 U(\nu_a)
\eeq
with (using the same sharp cut-off as in~\cite{Barnard:2013zea} for simplicity)
\beqs
    V(\mu) &=&  \frac{3}{2k} \mu^2 - \frac{3}{8\pi^2}\left( \Lambda^2 \mu^2 + \Lambda^4 \log\frac{\Lambda^2+\mu^2}{\Lambda^2} +
    \mu^4 \log\frac{\mu^2}{\Lambda^2+\mu^2} \right)\nonumber\\
    U(\nu) &=&  \frac{9}{k'} \nu^2 - \frac{1}{2\pi^2}\left( \Lambda^2 \nu^2 + \Lambda^4 \log\frac{\Lambda^2+\nu^2}{\Lambda^2} +
    \nu^4 \log\frac{\nu^2}{\Lambda^2+\nu^2} \right)
\eeqs

\begin{figure}
\centering
\includegraphics[width=.5\textwidth]{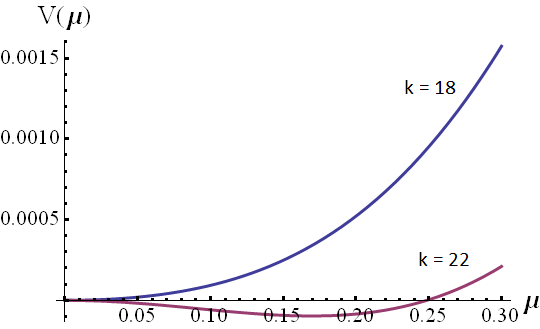}
\caption{\small Plot of the potential $V(\mu)$ in units $\Lambda=1$ for $k=18$ and $k=22$. The transition occurs at $k_{\mathrm{min}} \approx 20.$ For $k > k_{\mathrm{max}} \approx 64$, the minimum is at a value comparable to the cutoff and the approximation breaks down.
  \label{NJLplot}}
\end{figure}

A plot of the potential $V(\mu)$ is shown in Fig.~\ref{NJLplot}. (The potential $U(\nu)$ is obviously similar.) For large enough values of $k$ the minimum is attained at non-zero $\mu$ and the symmetry is broken. This is not a proof of symmetry breaking since we have no control on the actual values of the effective couplings. It does however point to the same direction as the previous argument and shows explicitly that, if symmetry breaking occurs, there is a basis in which $\langle \epsilon_{mnpq} \psi^{Imn}\psi^{Jpq} \rangle \propto \delta^{IJ}$, breaking $SU(5)\to SO(5)$, and
$ \tilde\chi_{m a'}\chi^{ma} \propto \delta_{a'}^a$, breaking $SU(3)\times SU(3)'\to SU(3)_c$. ($U(1)'$ is also broken while $U(1)_X$ is left unbroken.)

The Maximally Attractive Channel hypotheses (MAC)~\cite{Raby:1979my} indicates that the breaking of $SU(5)$ occurs at a higher scale compared to that of $SU(3)\times SU(3)'$. We can quantify the ratio of scales by a naive one-loop matching.

For the condensation in the $\psi\psi$ channel, the MAC is
$\mathbf{6}\times\mathbf{6} \to \mathbf{1}$, with attractive strength $r_{\psi\psi}=C(\mathbf{1})- C(\mathbf{6})-C(\mathbf{6})=-5$. ($C(\mathbf{R})$ is the quadratic Casimir of the irrep $\mathbf{R}$.) In the $\tilde\chi\chi$ channel we have a MAC $\bar{\mathbf{4}}\times\mathbf{4} \to \mathbf{1}$ and strength $C(\mathbf{1})- C(\bar{\mathbf{4}})-C(\mathbf{4})=-15/4$. (The chiral channels like $\psi\chi$ are always less attractive than both of the above.)

The one loop $SU(4)$ $\beta$-function coefficient with the $\psi$ removed is $b= -38/3$, having defined $(\mu \mathrm{d}/\mathrm{d}\mu) \alpha_{\mathrm HC} = b\,\alpha^2_{\mathrm HC}/2\pi$. Assuming that the condensates form when $|r| \alpha_{\mathrm HC} \approx 1$, we can relate the scales as
\beq
      \frac{\Lambda_{\psi\psi}}{\Lambda_{\tilde\chi\chi}} \approx \exp\left\{\frac{2\pi}{|b|}(|r_{\psi\psi}|-|r_{\tilde\chi\chi}|)\right\} \approx 1.9
\eeq
Again, none of these arguments is rigorous (see e.g.~\cite{Georgi:1981mh,Eichten:1981mu}) but it seems safe to assume that the $SU(5)$ breaking occurs at a higher scale. We shall be mostly concentrating on the
$SU(5)/SO(5)$ part, since this is where the EW dynamics resides. The effect of the remaining $SU(3)\times SU(3)'/SU(3)_c$ is just that of generating a color octet pNGB that couples in the obvious way. We denote by $f$ and $f'$ the respective decay constants.

\subsection{Running of the SM couplings}
\begin{figure}[t]
 \center
  \includegraphics[width=.5\textwidth]{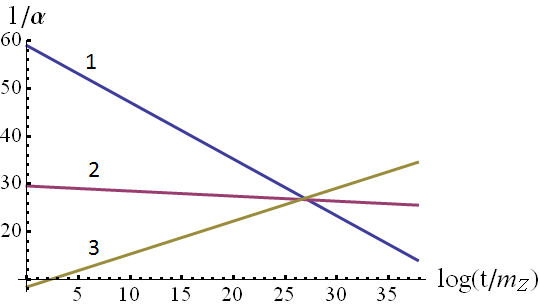}
 \caption{\small Running of the couplings in the present model. The matching is amusing but should not be taken seriously as it stands since it corresponds to a scale of $4.4~10^{13}$~GeV.}
\label{unification}
\end{figure}

Having a candidate UV theory at one disposal can also be used to analyze the impact of the extra matter fields on the unification of the SM coupling.
We should not expect any exact matching, since we have introduced a new gauge group and the new fields do not form complete multiplets. Morover, there is clearly some UV physics at higher scales still missing in order to explain the origin of the couplings between the hyperfermions and the SM fermions. At least though, one should check that the picture is not completely distorted, e.g. by the existence of Landau poles at low energies. In Fig.~\ref{unification} we present the one-loop running of the SM couplings $\alpha_3\equiv\alpha_s$, $\alpha_2\equiv\alpha_W$ and $\alpha_1\equiv 5 \alpha_Y /3$ for our model. The running is given by the equation
\beq
      \frac{{\mathrm d}}{{\mathrm d} t} \alpha_i^{-1} = - \frac{b_i}{2\pi}, \quad\hbox{with }~t=\log(\mu/m_Z)
\eeq
with~\footnote{For comparison, we recall the well known results $(b_1, b_2, b_3) = (41/10, -19/6, -7)$ and $(b_1, b_2, b_3) =(33/5, 1, -3)$ for the SM and MSSM respectively as well as
$(b_1, b_2, b_3) =(152/15, -2, -11/3)$ for the model~\cite{Barnard:2013zea}. The Reduced Planck scale corresponds to $t= 37.8$.}
\beq
      (b_1, b_2, b_3) = (112/15, 2/3, -13/3).
\eeq

It is amusing to see that the matching is improved, but this should not be taken seriously since the scale of the matching is way to small ($4.4~10^{13}$~GeV) for proton decay to be ignored. Perhaps the addition of the extra heavy fields that are necessary anyway to generate the four-fermi interactions could cure that. We checked some of the remaining models classified in~\cite{Ferretti:2013kya} and did not find any sign of unification. In fact, many suffer of problems from Landau poles.

\section{The IR theory}

Having discussed the basic features of the UV theory, we now present the IR effective theory. The two sets of fields that we will retain are the pNGBs and the top quark partners.

As before, we denote by $\Lambda$ the $SU(4)$ strong coupling scale, or, equivalently, the mass scale of a \emph{typical} composite state (i.e. neither Goldstone not the lightest top-like partner, that will be assumed to have lower mass $M$). $f$ is the $SU(5)/SO(5)$ pNGB decay constant. The ratio between $\Lambda$ and $f$ can be estimated as $\Lambda/f \equiv g \approx 4\pi/\sqrt{\NHC} = 2\pi$.

For guidance, a tuning parameter $\xi \equiv v^2/f^2 \approx  0.1$ gives $f \approx 800$~GeV and $\Lambda \approx 5$~TeV. The top-partner mass $M$ will lie somewhere in between $f$ and $\Lambda$. The UV description in terms of the $SU(4)$ gauge theory coupled to the SM is assumed to be valid up to a UV scale $\Lambda_{\mathrm UV} \gg \Lambda$ where the four-fermion interactions are generated.
We will not attempt to discuss the physics involved at $\Lambda_{\mathrm UV}$, but FCNC indicate that $\Lambda_{\mathrm UV} > 10^7$~GeV. 

\subsection{Composite fields}
\label{compositefields}

As far as the pNGBs are concerned, we argued in the previous section that the symmetry breaking pattern takes the form
\beqs
      \frac{\GF}{\HF} &=& \frac{SU(5) \times SU(3) \times SU(3)' \times U(1)_X \times U(1)' }{SO(5) \times SU(3)_c \times U(1)_X } \nonumber\\
          &=& \left( \frac{SU(5)}{SO(5)} \right) \times \left( \frac{SU(3) \times SU(3)'}{SU(3)_c} \right) \times  U(1)'
\eeqs

So far, all these bosons are massless and we now need to discuss how the coupling to the SM fields affects their spectrum.

The Goldstone boson $\eta'$ corresponding to $U(1)'$ is SM-neutral and will remain massless in our approximation. We will drop it from now on, but its role should be discussed in the cosmological context.

The EW breaking will be driven by the dynamics of the $SU(5)/SO(5)$ coset and for this we need to specify the embedding of the EW part of the SM gauge group $SU(2)_L\times U(1)_Y$ into $\HF$. This is done by first decomposing $SO(5) \to SO(4) \equiv SU(2)_L \times SU(2)_R$, then identifying a $U(1)_R$ subgroup of $SU(2)_R$ generated by $T_R^3$ and, lastly, setting $Y = T_R^3 + X$.

We take the vev for the $\psi$ bilinear $\langle \psi^I \psi^J\rangle $ proportional to $\delta^{IJ}$~\footnote{This is different from the most commonly used convention established in~\cite{Dugan:1984hq, Georgi:1984af, Georgi:1985nv}, where this coset was originally presented. The physical couplings are of course independent on the explicit representation chosen.}.
The 24 traceless hermitian generators of the fundamental irrep of $SU(5)$ are then decomposed into 10 imaginary anti-symmetric generators of $SO(5)$, generically denoted by $T^a$, and the remaining 14 traceless real symmetric broken generators, generically denoted by $S^i$, corresponding each to one Goldstone boson. The generators of $SO(4)$ are embedded into $SO(5)$ by padding the last row and column with zeros and choosing the remaining $4\times 4$ representation as in~\cite{DeSimone:2012fs}\footnote{Much of the notation in this work is influenced by this paper.}. It is convenient to have the expression for the generators of the $SU(2)_L \times U(1)_R$ subgroup of $SO(5)$:
\beqs
    &&T_L^1 =\frac{i}{2}
\begin{pmatrix}
 0 & 0 & 0 & -1 & 0 \\
 0 & 0 & -1 & 0 & 0 \\
 0 & 1 & 0 & 0 & 0 \\
 1 & 0 & 0 & 0 & 0 \\
 0 & 0 & 0 & 0 & 0
\end{pmatrix}, \qquad
T_L^2 =  \frac{i}{2}
\begin{pmatrix}
 0 & 0 & 1& 0 & 0 \\
 0 & 0 & 0 & -1 & 0 \\
 -1 & 0 & 0 & 0 & 0 \\
 0 & 1 & 0 & 0 & 0 \\
 0 & 0 & 0 & 0 & 0
\end{pmatrix}\nonumber\\
&&
T_L^3 =\frac{i}{2}
\begin{pmatrix}
 0 & -1 & 0 & 0 & 0 \\
 1 & 0 & 0 & 0 & 0 \\
 0 & 0 & 0 & -1 & 0 \\
 0 & 0 & 1 & 0 & 0 \\
 0 & 0 & 0 & 0 & 0
\end{pmatrix}, \qquad
T_R^3 = \frac{i}{2}
\begin{pmatrix}
 0 & -1 & 0 & 0 & 0 \\
 1 & 0 & 0 & 0 & 0 \\
 0 & 0 & 0 & 1 & 0 \\
 0 & 0 & -1 & 0 & 0 \\
 0 & 0 & 0 & 0 & 0
\end{pmatrix}
\eeqs

As far as the broken generators are concerned, we describe them by giving the explicit matrix for the Goldstone fields.
Decomposing the $SU(5)/SO(5)$ coset according to $SU(2)_L \times U(1)_R$ one finds~\cite{Dugan:1984hq} one totally SM neutral real boson $\eta$ (that will also be dropped in the following), one ``true Higgs" doublet $H$, a $Y$-neutral, $SU(2)_L$-triplet $\Phi_0$ and a charged one $\Phi_\pm$:
\beq
    {\mathbf{14}} \to {\mathbf 1}_{0} + {\mathbf 2}_{\pm 1/2} + {\mathbf 3}_{0} + {\mathbf 3}_{\pm 1} \equiv (\eta, H, \Phi_0, \Phi_\pm)
\eeq
For the Higgs, we will use the standard notation $H=(H_+, H_0)$, the indices denoting directly the electric charge
$Q = T_L^3 + Y \equiv T_L^3 + T_R^3$ (recall that all Goldstone bosons have $X=0$). For the triplets we use the notation
$\phi_{m_R}^{m_L}$, with $m_{R/L} = -1, 0, +1$ eigenvalues of $T^3_{R/L}$, e.g. $\Phi_0 \supset (\phi_0^{-}, \phi_0^{0}, \phi_0^{+})$,
and $\Phi_+ \supset (\phi_+^{-}, \phi_+^{0}, \phi_+^{+})$.
The electric charge is $Q=m_L+m_R$ and $(\phi_{m_R}^{m_L})^\dagger = \phi_{-m_R}^{-m_L}$.
There is thus~\cite{Georgi:1985nv} one double-charge scalar ($\phi_+^+$ and h.c.), two single-charge ones ($\phi_+^0$, $\phi_0^+$ and h.c.) and three neutral ones, (having dropped $\eta$ and $\eta'$), ($\phi_0^0$, $\Re\phi_+^-$ and $\Im\phi_+^-$). All these pNGB fit into the $SU(5)/SO(5)$ generators as

\beqs
    H &=&
        \begin{pmatrix}
          0 & 0 & 0 & 0 & -i H_+/\sqrt{2} \\
          0 & 0 & 0 & 0 & H_+/\sqrt{2} \\
          0 & 0 & 0 & 0 & i H_0/\sqrt{2} \\
          0 & 0 & 0 & 0 & H_0/\sqrt{2} \\
          -i H_+/\sqrt{2} & H_+/\sqrt{2} & i H_0/\sqrt{2} & H_0/\sqrt{2} & 0\\
        \end{pmatrix}\nn\\ \nn\\
   \Phi_0 &=& \begin{pmatrix}
                \phi_0^0/\sqrt{2} & 0 & i(\phi_0^- - \phi_0^+)/2 & (\phi_0^- + \phi_0^+)/2 & 0 \\
                0 & \phi_0^0/\sqrt{2} & (\phi_0^- + \phi_0^+)/2  & -i(\phi_0^- - \phi_0^+)/2  & 0 \\
                i(\phi_0^- - \phi_0^+)/2 & (\phi_0^- + \phi_0^+)/2 & -\phi_0^0/\sqrt{2} & 0 & 0 \\
                (\phi_0^- + \phi_0^+)/2 & -i(\phi_0^- - \phi_0^+)/2 & 0 & -\phi_0^0/\sqrt{2} & 0 \\
                0 & 0 & 0 & 0 & 0
              \end{pmatrix}\nonumber \\ \nn\\
    \Phi_+ &=& \begin{pmatrix}
                   \phi_+^+/\sqrt{2} & i\phi_+^+/\sqrt{2} & i\phi_+^0/2 & \phi_+^0/2 & 0 \\
                   i\phi_+^+/\sqrt{2} & -\phi_+^+/\sqrt{2} & -\phi_+^0/2 & i\phi_+^0/2 & 0 \\
                   i\phi_+^0/2 & -\phi_+^0/2 & \phi_+^-/\sqrt{2} & -i\phi_+^-/\sqrt{2}  & 0 \\
                   \phi_+^0/2 &  i\phi_+^0/2 & -i\phi_+^-/\sqrt{2}  & -\phi_+^-/\sqrt{2}  & 0 \\
                   0 & 0 & 0 & 0 & 0 \\
                 \end{pmatrix}
\eeqs

We combine these bosons as
\beq
    \Pi = H + H^\dagger + \Phi_0 + \Phi_+ + \Phi_+^\dagger, \quad\hbox{and}\quad
    \Sigma = \exp\left( \frac{i \Pi}{f}\right)  \label{eqf}
\eeq
Note that $\Pi$ is a real and symmetric matrix.
We will later argue that EW breaking takes place as expected, namely by giving a vev to the neutral component of $H$, $H_0 = h/\sqrt 2$. The remaining components of $H$ are then the true Goldstone bosons to be eaten by the $W^\pm$ and $Z$ and can be set to zero in the unitary gauge\footnote{Throughout the paper, we use the normalization where the $W$ mass, the vev of the canonically normalized $h$, and the decay constant $f$ are related by $m_{\mathrm{w}} =(g/2) f\sin(\langle h\rangle/f)$, i.e. \emph{the same} relation as used in the smaller coset $SO(5)/SO(4)$, yielding $v =f\sin(\langle h\rangle/f)=246$~GeV. We find this uniform definition less confusing than the one more commonly used for this coset, where $f$ is scaled by a factor 2.}. It is convenient to express the exponential exactly in $h$ and expand around the other fields, if necessary, using
\beq
     \Sigma = \Sigma(h) + \frac{i}{f} \int_0^1 {\mathrm d} s \Sigma((1-s)h) \left( \Phi_0 + \Phi_+ + \Phi_+^\dagger\right) \Sigma(s h) + \dots \label{expa}
\eeq
where, defining $\ch = \cos(h/f)$ and $\sh = \sin(h/f)$,
\beq
      \Sigma(h) =
            \begin{pmatrix}
              1 & 0 & 0 & 0 & 0 \\
              0 & 1 & 0 & 0 & 0 \\
              0 & 0 & 1 & 0 & 0 \\
              0 & 0 & 0 & \ch & i \sh \\
              0 & 0 & 0 & i \sh & \ch  \\
            \end{pmatrix}  \label{truehiggssigma}
\eeq

The Goldstone bosons in the $SU(3) \times SU(3)'/SU(3)_c$ coset transform in the $\mathbf 8$ of color. We simply write them as $\pi = \pi^a \lambda^a/2$ where $\lambda^a$ are the usual Gell-Mann matrices.

Moving on to the top quark partners, one of the advantages of having a candidate UV completion is that it allows one to motivate picking a particular irrep of $\HF$ for such objects. We collect in Table~\ref{IRspinhalf} all the spin-half $SU(3)_c$ triplet excitations created by the composite states obtained from Table~\ref{compositefermions}, now classified according to the unbroken global symmetry.

\begin{table}
  \centering
  \begin{tabular}{c|c|l|}
    % after \\: \hline or \cline{col1-col2} \cline{col3-col4} ...
       Object &$SO(5) \times SU(3)_c\times U(1)_X $\\
      \hline
      $\tilde\chi\psi\tilde\chi$,~~~ $\bar\chi\psi\bar\chi$, ~~~$2 \times \bar{\chi}\bar\psi\tilde\chi$
                                  & \hspace{0.8cm} $(\mathbf{5}, \mathbf{3})_{2/3\phantom{-}}$ \\

      $\chi\psi\chi$,~~~ $\bar{\tilde\chi}\psi\bar{\tilde\chi}$, ~~~$2 \times \bar{\tilde\chi}\bar\psi\chi$
                                  & \hspace{0.8cm} $(\mathbf{5}, \bar{\mathbf{3}})_{-2/3}$ \\
    \hline
  \end{tabular}
  \caption{\small The spin $1/2$ color triplets particles created by composite fermions after symmetry breaking. Shown are the LH combinations. The charge conjugates of the operators in the second row combine with the ones in the first row to give a total of four Dirac spinors. We assume without proof that one of them is significantly lighter than the others. Similar considerations can be made for the sextets and the spin $3/2$ resonances, although in this case we don't need to assume that some of them are lighter than the generic scale $\Lambda$.}\label{IRspinhalf}
\end{table}

We identify both the $\mathbf 5$ and  $\bar{\mathbf 5}$ of $SU(5)$ with the $\mathbf 5$ of $SO(5)$ and construct the $SU(3)_c$ irreps from
$SU(3) \times SU(3)'$ using ${\mathbf 3}\times {\mathbf 3} = \bar{\mathbf 3} + {\mathbf 6}$.
We do not consider any longer the spin 3/2 objects nor the color sextets, that we assume correspond to heavier states at the scale $\Lambda$. These states however are additional prediction of this UV completion and would allow one to discern it from other possibilities if experiments were performed at a higher scale.

The breaking of the global symmetry is what allows us to form Dirac fields out of the LH objects displayed in Table~\ref{IRspinhalf} and their RH conjugates.

One assumption (that we will not attempt to justify in this work) is that one linear combination of operators creates a fermionic resonance of mass $M$ that is lighter than the generic resonance scale $\Lambda$. This is not too unreasonable since we are asking for less that a factor ten suppression. Thus, we will assume the existence of \emph{one Dirac field} $\Psi$, of mass $M$, transforming in the $(\mathbf{5}, \mathbf{3})_{2/3}$ of $\HF$.

To extract the EW quantum numbers for these fields, note that
\beq
\begin{array}{cc}
     SO(5) \times SU(3)_c \times U(1)_X & (\mathbf{5}, \mathbf{3})_{2/3} \\
            \downarrow & \downarrow \\
   \GCUS \equiv SU(3)_c \times SU(2)_L \times SU(2)_R \times U(1)_X & (\mathbf{3},\mathbf{2},\mathbf{2})_{2/3}+(\mathbf{3},\mathbf{1},\mathbf{1})_{2/3}\\
            \downarrow & \downarrow \\
   \GSM \equiv SU(3)_c \times SU(2)_L \times U(1)_Y & (\mathbf{3},\mathbf{2})_{7/6} + (\mathbf{3},\mathbf{2})_{1/6} + (\mathbf{3},\mathbf{1})_{2/3}\\
            \downarrow & \downarrow \\
    SU(3)_c \times U(1)_{\mathrm{e.m.}} & \mathbf{3}_{5/3} + 3 \times  \mathbf{3}_{2/3} +  \mathbf{3}_{-1/3}
\end{array}
\eeq

We have thus succeeded in contracting a partner for the LH SM field $q_L^3 = (t_L, b_L)$, namely the RH projection of the Dirac field
$(T, B) \in (\mathbf{3},\mathbf{2})_{1/6}$ above, and a partner to the RH SM field $t_R$, namely the LH projection of the Dirac field
$R \in (\mathbf{3},\mathbf{1})_{2/3}$. We will only consider the mixing between the composite fermions and the third family.

Notice that we do not find partners to the remaining SM fields, including $b_R$. We could simply ignore this problem by focusing on the more pressing issue of the top mass, but we argue that for the remaining particles it is still feasible to consider a bilinear mass term as in the early constructions~\cite{Kaplan:1983fs} and we will do so in the following. Given the quantitative difference of the top quark mass and the difficulty in finding an acceptable model giving all partners, this option seems more attractive to us.

We have already discussed the components $T, B$ and $R$. Denoting the remaining fields by $(X, Y)\in (\mathbf{3},\mathbf{2})_{7/6}$, we write the
full $(\mathbf{5}, \mathbf{3})_{2/3}$ multiplet as
\beq
     \Psi = \frac{1}{\sqrt 2}
              \begin{pmatrix}
                i B - i X \\
                 B + X \\
                i T + i Y \\
                - T + Y \\
                \sqrt{2} i R \\
              \end{pmatrix}
\eeq
After EW symmetry breaking, the fields $T$, $Y$ and $R$ acquire electric charge $2/3$ and mix with the top quark. Similarly, the field $B$ acquires an electric charge $-1/3$ and mixes with the bottom quark. The field $X$ has charge $5/3$ and is a generic prediction of many models of this type.

\subsection{EW symmetry breaking}

The most pressing issue is to show that the desired misalignment of the Higgs field $H$, leading to the correct EW symmetry breaking, can occur.

Precise quantitative computations are precluded by our lack of control of the strong dynamics. What we can hope to do is to show that the couplings of the SM fields to the pNGBs are such that the misalignment can occur for the Higgs doublet $H$ but not for the other fields.
We will consider top-quark-driven misalignment as proposed in~\cite{Agashe:2004rs}.

We want to write an effective action coupling the pNGBs to the SM vector bosons and fermions.
Under a generic global $g \in SU(5)$ transformation, $\Sigma$ in (\ref{eqf}) transforms non-linearly as
$\Sigma \to g \Sigma h^T(\Pi, g)$ with $h(\Pi, g)\in SO(5)$, a real matrix. In this case, we are allowed to construct a simpler object
$U = \Sigma\Sigma^T \equiv \exp\left( \frac{ 2 i }{f}\Pi\right)$ transforming linearly as $U \to g U g^T \equiv {\mathrm{Symm}}_g \circ U$.

The couplings to the vector bosons are obtained from
\beq
      {\mathcal L} \supset \frac{f^2}{16} \tr \left( (D_\mu U)^\dagger  D^\mu U \right) \label{DUDU}
\eeq
where,
\beq
    D_\mu U = \partial_\mu U - i g W^a_\mu [T_L^a,\, U] - i g' B_\mu[T_R^3,\, U].
\eeq
For simplicity, we will only consider contributions from the $SU(2)_L$ EW bosons $W^a_\mu$.

To couple the pNGBs to the SM fermions we need to determine the spurionic embeddings\footnote{For a given SM field $q$, we denote by $\hat q$ a field with the same dynamical content but transforming as a full multiplet of $\GF$. For conciseness, we call the whole $\hat q$ ``spurion'' and never write down the auxiliary fields.} by considering how they can be coupled to the composite field $\Psi$.
Given $\Psi$ in the $\mathbf 5$ of $SO(5)$ as above, we can construct the operators $\Sigma \Psi$ and $\Sigma^* \Psi$ transforming in the $\mathbf 5$ and $\bar{\mathbf 5}$ of $SU(5)$ respectively. This fact forces us to pick, as spurionic embedding of the elementary quarks $q_L$ and $t_R$, the $\mathbf 5$ and $\bar{\mathbf 5}$ representation as well. We write
\beq
     \hat{q}_L = \frac{1}{\sqrt 2}
              \begin{pmatrix}
                i b_L \\
                 b_L \\
                i t_L \\
                - t_L \\
               0 \\
              \end{pmatrix}, \quad\hbox{and}\quad
     \hat{t}_R =
              \begin{pmatrix}
                0 \\
                0 \\
                0 \\
                0 \\
              i t_R \\
              \end{pmatrix}
\eeq
The coupling with the $\Psi$ will be important later, now we focus on the induced terms. In momentum space they read
\beq
     {\mathcal L} \supset G(p)\left( \bar{\hat q}_L U \hat t_R + \bar{\hat t}_R U^* \hat{q}_L \right)
\eeq
where $G(p)$ is a form factor depending on the strong dynamics and the rest of the fields is evaluated at zero momentum.
Notice that kinetic terms of type $\bar{\hat q}_L U \!\not\! p \hat{q}_L$ are \emph{not} allowed since $U$ is in the ${\mathbf{15}}$.
For the same reason, we cannot pick both the spurions in the ${\mathbf{5}}$.

We start by expanding around the \emph{unbroken} vacuum $\Pi=0$ and look for possible destabilizing effects.
Once we convince ourselves that the breaking occurs when $H_0$ gets a vev, we set all other fields to zero and treat $H_0$ to all orders.

\begin{figure}[t]
 \center
 \subfigure[Contribution of the $SU(2)_L$ gauge bosons]{%
  \includegraphics[width=.40\textwidth]{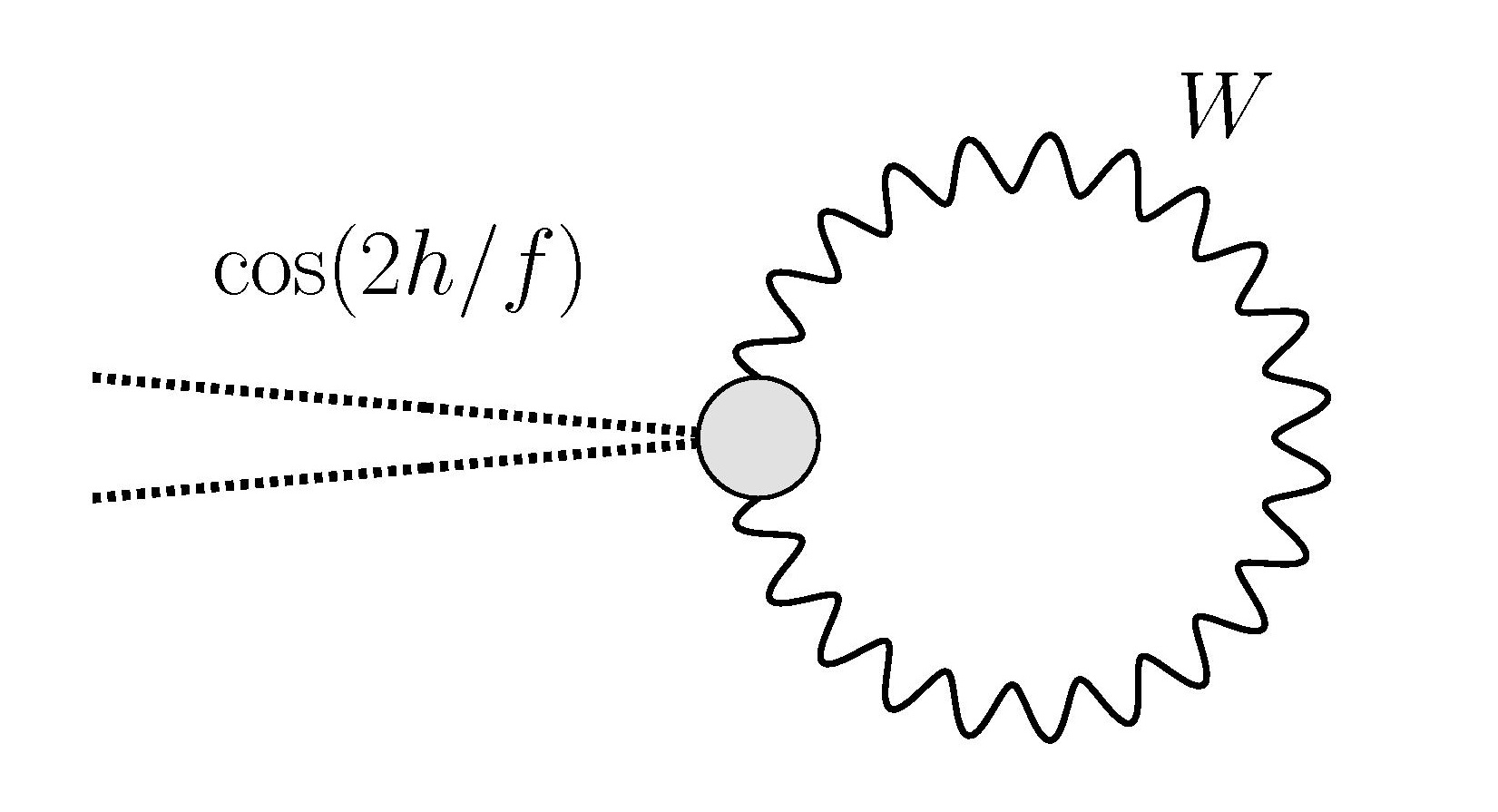}}
 \subfigure[Contribution of the top quark]{%
  \includegraphics[width=.48\textwidth]{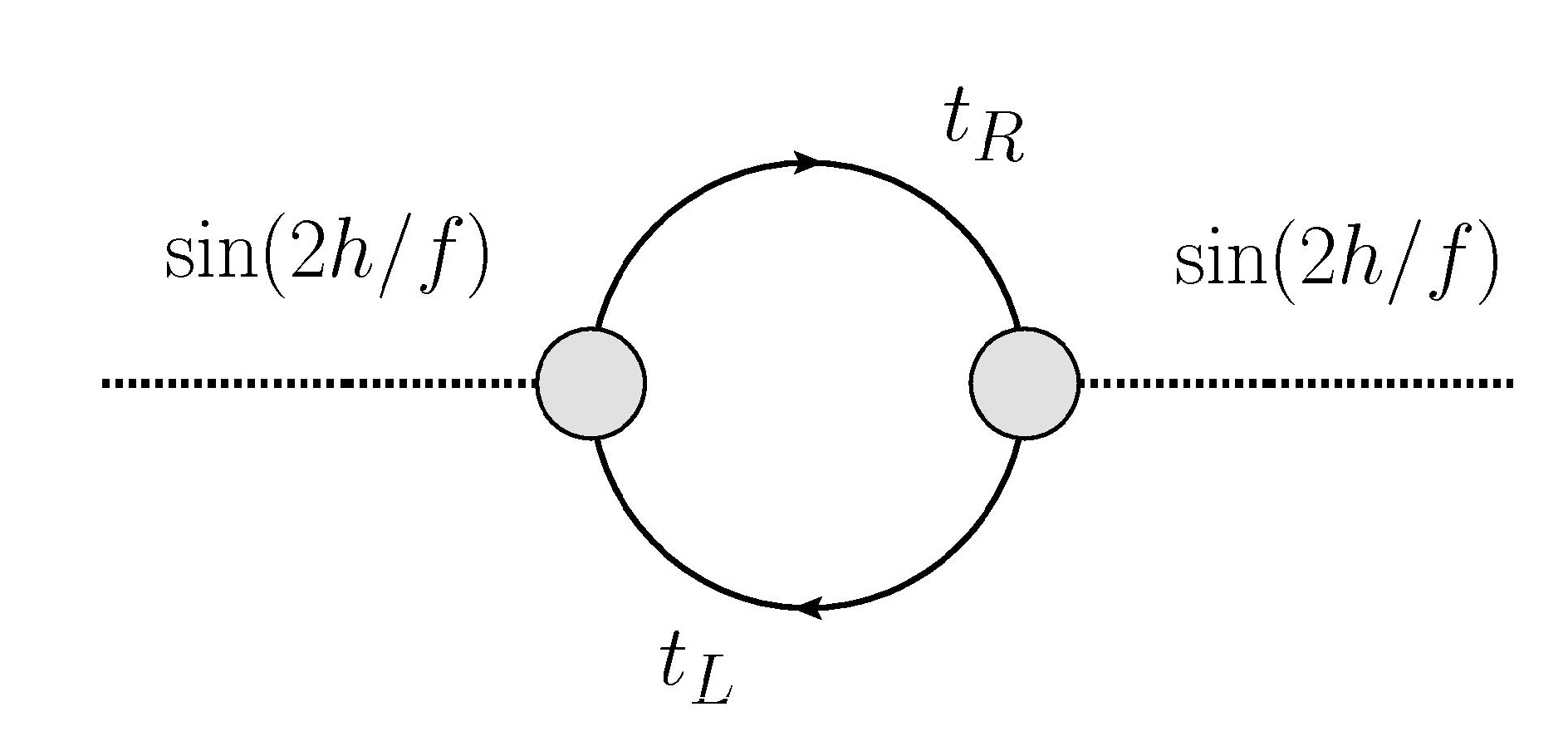}}
 \caption{\small The leading order contributions to the induced Higgs potential.}
\label{CWpot}
\end{figure}

The couplings in (\ref{DUDU}) with the $SU(2)_L$ EW bosons is proportional to
\beq
    \tr(T^a_L T^a_L\, \Pi\, \Pi - T^a_L\, \Pi\, T^a_L \,\Pi ) = \frac{3}{2} H^\dagger H + 4 \Phi_+^\dagger \Phi_+ + 2 \Phi_0^\dagger \Phi_0 \label{titi}
\eeq
Vector couplings of this type do not misalign the vacuum~\cite{Witten:1983ut}.
This means that they will contribute to the pNGB potential with a positive overall coefficient to the combination in~(\ref{titi}).
The only possible negative contributions must come from the fermionic couplings, which are proportional to
\beq
      \bar{\hat q}_L \Pi \hat t_R + \bar{\hat t}_R \Pi \hat{q}_L = \frac{2}{f} (\bar{q}_L H^\dagger t_R - \bar{t}_R H q_L).
\eeq
Hence, it is only for the field $H$ that we can expect a misalignment. We now set $H_0 = h/\sqrt 2$, all other fields to zero, and write
\beq
       U(h) = \Sigma(h)\Sigma(h)^T \equiv \Sigma(2h)  \label{truehiggsU}
\eeq
yielding
\beqs
    && W^a_\mu W^b_\mu \tr(U(h) T_L^a U(h)^\dagger T_L^b) = \frac{1}{2}\left(1 + \cos(2h/f)\right) W_\mu^c W_\mu^c
         \nn\\
    && \bar{\hat q}_L U(h) \hat t_R + \bar{\hat t}_R U(h)^* \hat{q}_L = \frac{1}{\sqrt 2} \sin(2h/f) (\bar{t}_L t_R + \bar{t}_R t_L).
\eeqs
The contribution to the Coleman-Weinberg potential~\cite{Coleman:1973jx} is given, to leading order, by the diagrams in Fig.~\ref{CWpot}.
We can then summarize the contribution of the integral over the resonances of the strong sector by two dimensionless numbers $\alpha$ and $\beta$ as done in e.g.~\cite{Contino:2010rs}
\beq
     V(h)\propto \alpha \cos (2h/f) - \beta \sin^2 (2h/f).
\eeq
An acceptable EW breaking minimum will be attained for $\beta \gtrsim |\alpha|/2$ at
$\sin^2 (2 \langle h \rangle/f) = 1 - (\alpha/2\beta)^2$. Recalling that with our conventions $v = f \sin(\langle h\rangle/f)$, we get a relation between the fine-tuning parameter $\xi$ and the terms in the Higgs potential
\beq
     \xi \equiv \left(\frac{v}{f}\right)^2\approx \frac{1}{4}\left( 1 - \left(\frac{\alpha}{2\beta}\right)^2 \right),
\eeq
i.e. a factor of four improvement over the minimal coset. We believe it makes sense to compare the two because the relation between $v$, $\langle h\rangle$ and $f$ has been chosen to be the same for both. To our knowledge, this last observation was first made in~\cite{Vecchi:2013bja}, but see e.g.~\cite{Katz:2005au, Gripaios:2009pe, Galloway:2010bp} for related recent work.

\begin{figure}[t]
 \center
  \includegraphics[width=.4\textwidth]{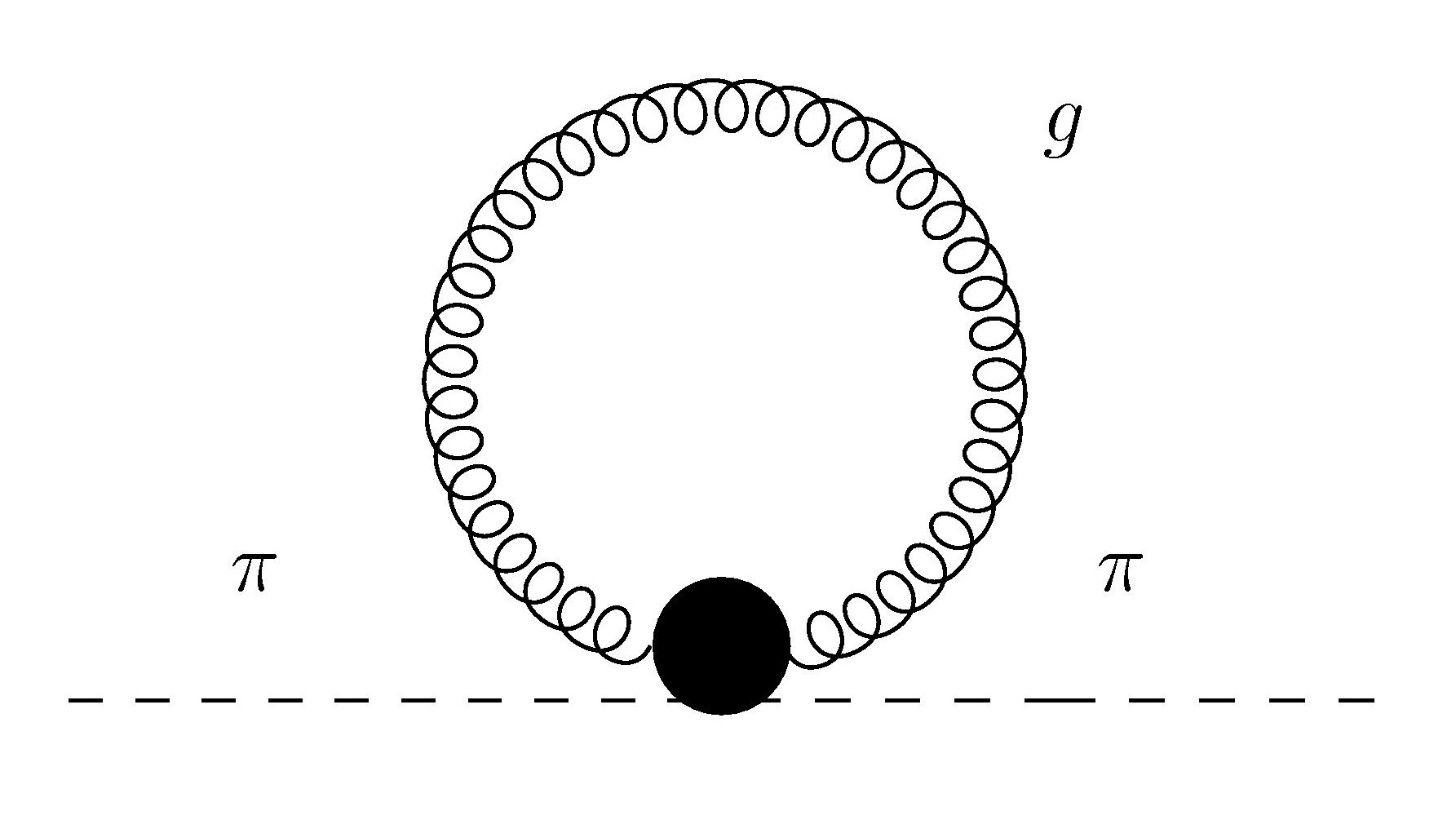}
 \caption{\small The gluon contribution to the mass of the pNGB $\pi$. }
\label{piloop}
\end{figure}

A simpler analysis can be done to show that the color octet $\pi^a$ will not be destabilized and thus color remains unbroken. These pNGBs will gain a mass that we can roughly estimate from the diagram in Fig.~\ref{piloop} as
\beq
      m_\pi^2 \approx \frac{3\times 8 \times \alpha_s}{4\pi} \Lambda^2 \approx ( 2~{\mathrm{TeV}} )^2 \label{pimass}
\eeq
In~(\ref{pimass}), $3$ is the number of off-shell gluon polarizations, $8$ the color factor and $\Lambda^2$ summarizes the effects of the strong sector resonances. One could be more precise and use sum-rules to write this contribution in terms of the masses of the lowest lying states but not much is gained in this case since we do not have experimental information on their mass. In this case there can also be contributions from the quarks kinetic terms but we expect them to be subleading compared to the gluons.

\subsection{The fermionic mass terms}

In subsection~\ref{compositefields} we discussed the transformation properties of the composite pNGB and fermions. We saw that, in order to couple linearly to the top quark, we needed to embed the elementary fields $q_L$ and $t_R$ into spurions in the $\mathbf 5$ and $\bar{\mathbf 5}$ of $SU(5)$ respectively. However, the lack of candidate partners for the remaining fermions made it impossible to construct all masses this way. The complexity (and unlikeliness) of UV completion providing all partners made us propose a compromise: The remaining masses are constructed with \emph{bilinears}. At least the fine tuning is mitigated by only needing to achieve a mass of the order of a few GeV. Even this requires some care though, since, in order to preserve the $SU(5)$ invariance, we need to pick the representation for the spurions in a way compatible with the choices in subsection~\ref{compositefields}.

Let us consider the quarks and denote by ${\hat q}^i_L$, ${\hat u}^i_R$, and ${\hat d}^i_R$, the $SU(5)$ spurionic embeddings, where $i=1,2,3$ is the family number. We have already encountered ${\hat q}_L$ (no index $i$) and ${\hat t}_R$ which have the same physical field content as  ${\hat q}^3_L$ and ${\hat u}^3_R$ but arranged in a different irrep.

Since the pNGB fields carry no $U(1)_X$ charge and the quarks mix, the remaining spurions must have the same $U(1)_X$ charge as the top partners. This fact, together with the usual relation between $X$, $Y$ and $T_R^3$, fixes the quantum numbers displayed in Table~\ref{spurionqn}.
\begin{table}
  \centering
\begin{tabular}{|c|c|c|c|c|}
  \hline
  % after \\: \hline or \cline{col1-col2} \cline{col3-col4} ...
   & $SU(3)_c$ & $U(1)_X$ &$ SU(2)_L$ & $U(1)_R$ \\
  ${\hat q}^i_L$ & $\mathbf 3$ & 2/3 & $\mathbf 2$ & -1/2 \\
  ${\hat u}^i_R$ & $\mathbf 3$ & 2/3 & $\mathbf 1$ & 0 \\
  ${\hat d}^i_R$ & $\mathbf 3$ & 2/3 & $\mathbf 1$ & -1 \\
  \hline
\end{tabular}
  \caption{\small The remaining spurions quantum numbers.}\label{spurionqn}
\end{table}
Now we want to embed the $ SU(2)_L \times U(1)_R$ irreps of Table~\ref{spurionqn} into irreps of $SU(5)$ that allow to construct mass terms $\bar{{\hat q}}^i_L U^* {\hat u}^j_R$ and $ \bar{{\hat q}}^i_L U {\hat d}^j_R $. One sees that, restricting to at most ``two-index" irreps, the only solution that allows reproducing Table~\ref{spurionqn} and constructing the needed mass terms is
${\hat{q}}^i_L \in {\mathbf{24}}$, ${\hat u}^i_R \in \mathbf{10}$, and ${\hat d}^i_R \in \bar{\mathbf{10}}$, the adjoint, anti-symmetric and conjugate anti-symmetric irreps respectively\footnote{The $\mathbf{10}$ and $\bar{\mathbf{10}}$ irreps can be used interchangeably. That the fundamental irrep cannot be used can be inferred by the need to reproduce $T^3_R=-1$ for ${\hat{d}}^3_R$. The spurions are normalized to have canonical kinetic energy, e.g.
$\tr(\bar{{\hat q}}^i_L \not\! \partial {{\hat q}}^i_L ) = \bar{{q}}^i_L \not\! \partial q^i_L$.}. Setting all the auxiliary fields to zero we have, explicitly

\beqs
   && {\hat q}^i_L = \frac{1}{2}
                     \begin{pmatrix}
                       0 & 0 & 0 & 0 & i d_L^i \\
                       0 & 0 & 0 & 0 & d_L^i \\
                       0 & 0 & 0 & 0 & i u_L^i \\
                       0 & 0 & 0 & 0 & - u_L^i \\
                       i d_L^i & d_L^i & i u_L^i & -u_L^i & 0 \\
                     \end{pmatrix}, ~~
    {\hat u}^i_R = \frac{1}{2}
                     \begin{pmatrix}
                       0 & u_R^i & 0 & 0 & 0 \\
                       - u_R^i & 0 & 0 & 0 & 0 \\
                       0 & 0 & 0 & - u_R^i & 0 \\
                       0 & 0 & u_R^i & 0 & 0 \\
                       0 & 0 & 0 & 0 & 0 \\
                     \end{pmatrix}, \nn\\\nn\\
   && {\hat d}^i_R = \frac{1}{2\sqrt 2}
                     \begin{pmatrix}
                       0 & 0 & i d_R^i & - d_R^i & 0 \\
                       0 & 0 & d_R^i & i d_R^i & 0 \\
                       - i d_R^i & - d_R^i & 0 & 0 & 0 \\
                         d_R^i & - i d_R^i & 0 & 0 & 0 \\
                       0 & 0 & 0 & 0 & 0 \\
                     \end{pmatrix}. \label{explicitspurion}
\eeqs

The same construction works for the leptons with the only difference that now the $U(1)_X$ charges are taken to be zero.

We are now ready to write down the mass terms for the top and bottom quarks, including the contribution to $b_R \in \hat{d}_R^3$. We concentrate on the third family and write
\beq
   {\mathcal L} \supset \frac{M}{2} \bar\Psi \Psi + \lambda_q f \bar{\hat{q}}_L \Sigma \Psi_R + \lambda_t f \bar{\hat{t}}_R \Sigma^* \Psi_L
       + \sqrt{2} \mu_b\, \tr\left(\bar{\hat{q}}_L^3 U{\hat d}^3_R \right)+\hc \label{masses}
\eeq
The dimensionless couplings $\lambda_q$ and $\lambda_t$ between SM fields and the composite fermion are expected to be of the same order and control the mass of the top quark. The mass parameter $\mu_b$ is required to give a mass to the bottom quark and we ignore the subleading
quadratic terms like $\bar{\hat q}_L U \hat t_R$ and $\tr\left(\bar{\hat{q}}_L^3 U^* {\hat u}^3_R \right)$ for the top.

Inserting~(\ref{truehiggsU}) and (\ref{explicitspurion}) in ~(\ref{masses}), we obtain the following mass matrices

\beq
        {\mathcal{M}}_T =
                \begin{pmatrix}
                0 & \frac{\lambda_q}{2}f(1+\ch) & \frac{\lambda_q}{2}f(1-\ch) & \frac{\lambda_q}{\sqrt 2} f\sh \\
                \frac{\lambda_t}{\sqrt 2}f\sh & M & 0 & 0 \\
                -\frac{\lambda_t}{\sqrt 2}f \sh & 0 & M & 0 \\
                \lambda_t f \ch & 0 & 0 & M \\
                \end{pmatrix}
\eeq
and
\beq
        {\mathcal{M}}_B =
                \begin{pmatrix}
               \mu_b \sh \ch & \lambda_q f \\
               0 & M \\
                \end{pmatrix}
\eeq
in terms of which the lagrangian~(\ref{masses}) can be written as
\beq
      {\mathcal L} \supset
                      (\bar{t}_L, \bar{T}_L, \bar{Y}_L, \bar{R}_L)\cdot {\mathcal{M}}_T \cdot
                      \begin{pmatrix}
                        t_R \\ T_R \\ Y_R \\ R_R
                      \end{pmatrix} +
                    (\bar{b}_L, \bar{B}_L)\cdot {\mathcal{M}}_B \cdot
                      \begin{pmatrix}
                        b_R \\ B_R
                      \end{pmatrix} + \hc
\eeq

The lowest singular values of the two mass matrices have to be made coincide with the known masses of the top and bottom quarks.
For the top sector, we expand the lowest mass to leading order in the higgs vev $v$ to find
\beq
     m_t \approx \frac{\sqrt{2} M f \lambda_q \lambda_t}{\sqrt{M^2 +\lambda_q^2 f^2}\,\sqrt{M^2 +\lambda_t^2 f^2}}\, v,
\eeq
proportional to the product $\lambda_q \lambda_t$, in agreement with diagrammatic expectations.
The remaining masses have non-vanishing values even for $v \to 0$ and are, to zeroth order,
equal to $M$, $\sqrt{M^2 + \lambda^2_q f^2}$ and $\sqrt{M^2 + \lambda^2_t f^2}$

For the bottom quark we can go to the mass eigenstates by writing
\beqs
    &&
      \begin{pmatrix}
        b_L \\ B_L
      \end{pmatrix} =
               \begin{pmatrix}
                 \cos\lambda & \sin\lambda \\
                 -\sin\lambda & \cos\lambda \\
               \end{pmatrix}
      \begin{pmatrix}
        b'_L \\ B'_L
      \end{pmatrix}
        \equiv R_\lambda
      \begin{pmatrix}
        b'_L \\ B'_L
      \end{pmatrix} \nn\\
     && \begin{pmatrix}
        b_R \\ B_R
      \end{pmatrix} =
               \begin{pmatrix}
                 \cos\rho & \sin\rho \\
                 -\sin\rho & \cos\rho\\
               \end{pmatrix}
      \begin{pmatrix}
        b'_R \\ B'_R
      \end{pmatrix}
         \equiv R_\rho
      \begin{pmatrix}
        b'_R \\ B'_R
      \end{pmatrix}
    \label{rot}
\eeqs
with
\beq
      \tan 2\rho = \frac{2 \mu_b f \lambda_q \sh \ch  }{M^2 + \lambda_q^2 f^2 - \mu_b^2  \sh^2 \ch^2},
      \quad\hbox{and}\quad
      \tan 2\lambda = \frac{2 \lambda_q f M}{M^2 - \lambda_q^2 f^2 -\mu_b^2 \sh^2 \ch^2}.
\eeq
The mass of the $b$ quark is, to lowest order in the Higgs vev,
\beq
       m_b \approx \frac{\mu_b M}{f\,\sqrt{M^2 + \lambda_q^2 f^2}}\, v \label{mb}
\eeq
and the remaining mass is equal, to zeroth order,  to $\sqrt{M^2 + \lambda^2_q f^2}$, thus nearly degenerate with one of the top partners.

\subsection{The fermionic currents}

\label{currcurr}
We now compute the contribution of the fermionic partners to the EW currents. For this, we need first to define the matrix-valued one-forms
\beq
     p_\mu = \sum_{i=1}^{14} S^i \tr\big(S^i\Sigma^{-1}(i \partial_\mu \Sigma + e V_\mu \Sigma)\big), \quad
     v_\mu = \sum_{a=1}^{10} T^a \tr\big(T^a\Sigma^{-1}(i \partial_\mu \Sigma + e V_\mu \Sigma)\big)
\eeq
where
\beq
       V_\mu = W^+_\mu\frac{T_L^1 + i T_L^2}{\sqrt2 \sw} +  W^-_\mu\frac{T_L^1 - i T_L^2}{\sqrt2 \sw}+
       (A_\mu + \frac{\cw}{\sw} Z_\mu)T_L^3 + (A_\mu - \frac{\sw}{\cw} Z_\mu)T_R^3
\eeq
is the matrix-valued SM gauge field and  $e$ is the electric coupling constant.

The gauge currents associated to the composite fermion can be read off from
\beqs
      {\mathcal L}& \supset & \bar\Psi\gamma^\mu(\frac{2}{3} e A_\mu {\mathbf 1}- \frac{2\sw}{3\cw} e Z_\mu {\mathbf 1} + v_\mu)\Psi +
      K \bar\Psi\gamma^\mu p_\mu \Psi \nn\\
      &&\equiv e (J_A^\mu A_\mu + J_Z^\mu Z_\mu + J_{W^+}^\mu W^+_\mu + J_{W^-}^\mu W^-_\mu)+\dots \label{lagri}
\eeqs
where the only undetermined constant is $K$ and the dots represent terms without gauge fields.
Extracting the coefficients, we obtain, restricting to the coupling with the pNGB $h$ only,

\beqs
     J_Z^\mu &=& C_{XX} \bar X \gamma^\mu X + C_{TT} \bar T \gamma^\mu T + C_{YY} \bar Y \gamma^\mu Y + C_{RR} \bar R \gamma^\mu R \nn\\
              && + C_{BB} \bar B \gamma^\mu B + C_{RT} (\bar R \gamma^\mu T + \hc) +
                   C_{RY} (\bar R \gamma^\mu Y + \hc) + C_{TY} (\bar T \gamma^\mu Y + \hc)\nn \\
     J_{W^+}^\mu &=& C_{XT} \bar X \gamma^\mu T + C_{XY} \bar X \gamma^\mu Y + C_{XR} \bar X \gamma^\mu R  \nn\\
                && + C_{TB} \bar T\gamma^\mu B + C_{YB} \bar Y \gamma^\mu B + C_{RB} \bar R \gamma^\mu B \label{currents}
\eeqs

and, of course, $J_{W^-}^\mu = (J_{W^+}^\mu)^\dagger$ and
\beq
    J_{A}^\mu = \frac{5}{3} \bar X \gamma^\mu X + \frac{2}{3} (\bar T \gamma^\mu T + \bar Y \gamma^\mu Y + \bar R \gamma^\mu R)
                - \frac{1}{3} \bar B \gamma^\mu B.
\eeq

The coefficients in~(\ref{currents}) are given by
\beqs
      C_{XX} &=& \frac{1}{\sw\cw}\left(\frac{1}{2} - \frac{5 \sw^2}{3}\right)\nn\\
      C_{TT} &=& -\frac{2\sw}{3\cw} + \frac{\ch}{2 \sw \cw}  \nn\\
      C_{YY} &=& -\frac{2\sw}{3\cw} - \frac{\ch}{2 \sw \cw}  \nn\\
      C_{RR} &=& -\frac{2\sw}{3\cw} \nn\\
      C_{BB} &=&  \frac{1}{\sw\cw}\left(-\frac{1}{2} + \frac{\sw^2}{3}\right) \nn\\
      C_{TY} &=& 0 \nn\\
      C_{RT} =  C_{RY}  &=& \frac{K\, \sh}{2 \sqrt2 \sw\cw} \nn\\
      C_{XT} = C_{YB} &=& \frac{1-\ch}{2 \sqrt2 \sw} \nn\\
      C_{XY} = C_{TB} &=& \frac{1+\ch}{2 \sqrt2 \sw}  \nn\\
      C_{RB} = - C_{XR} &=& \frac{K\, \sh}{2 \sw}  \label{Ccoeff}
\eeqs

The important point to notice is the value of $ e C_{BB}$, which coincides with the analogous coefficient arising from the elementary $b_L$. This guarantees that, after rotating to the mass eigenbasis with the matrices~(\ref{rot}), the branching ratio $Z\to b \bar b$ does not suffer large corrections.
This is an explicit realization of the mechanism described in~\cite{Agashe:2006at}. Here the situation is satisfactory since the UV completion has \emph{forced us} to choose a $b_L$ belonging to one of the ``custodial irreps'' described in~\cite{Agashe:2006at}.

\subsection{Additional couplings}

There are infinite series of additional couplings dictated by the non-linear structure of the Lagrangian.
For instance, so far we have not considered the colored pNGBs, but their interactions can be written down in analogy with three-flavor QCD, with the difference that now the ``baryons'' $\Psi$ are in a triplet
\beq
    {\mathcal{L}} \supset \frac{1}{2}\bar\Psi\gamma^\mu\left( g_s G_\mu^a + \frac{\lambda_A}{f'}\gamma^5 \partial_\mu \pi^a +
    \frac{1}{f'^2} f^{abc} \pi^b \partial_\mu \pi^c + \dots\right)\lambda^a\Psi.
\eeq
(Here $G_\mu^a$ is the gluon, $g_s$ the QCD coupling constant and $\lambda_A$ the analog of the axial coupling.)

The term in~(\ref{lagri}) containing $p_\mu$ also gives rise to derivative interactions with the Higgs field of type
\beq
   {\mathcal{L}} \supset \frac{i K}{\sqrt 2 f}(\bar R\gamma^\mu Y - \bar R\gamma^\mu T)\partial_\mu h + \hc
\eeq
as well as couplings with the other pNGBs. Among these, there are non-derivative $1/f$-suppressed couplings between two composite fermions, a gauge field and a pNGB that could also be relevant for phenomenology.

Lastly, the mixing between composite and elementary quarks in (\ref{masses}) also gives rise to couplings with the additional pNGBs. Here we present only those that survive the limit $\langle h\rangle/f \to 0$
\beqs
       {\mathcal{L}} &\supset& \lambda_q\left(\bar b_L Y_R\phi_-^0 - \bar t_L X_R \phi_-^0
        - i \sqrt{2} \bar b_L X_R \phi_-^- + i \sqrt{2}\bar t_L Y_R \phi_-^+ + \frac{i}{\sqrt{2}}\bar b_L B_R \phi_0^0\right. \nn \\
         && \left.- \frac{i}{\sqrt{2}}\bar t_L T_R \phi_0^0
        - \bar b_L T_R \phi_0^- + \bar t_L B_R \phi_0^+ \right) + \hc
\eeqs
with no term arising in this limit from the couplings to the $t_R$.

\section{Discussion}

We presented a model of partial compositeness motivated by an UV completion based on a $SU(4)$ gauge group. This group was shown
in~\cite{Ferretti:2013kya} to be the only unitary group allowing for custodial symmetry and top partners while retaining asymptotic freedom. It is also the preferred one if one wants to avoid scalar color triplets and sextets.

The fields in the IR theory carrying SM charges consists of the standard $SU(5)/SO(5)$ pNGB coset, a color octet pNGB, one fermion mixing with the bottom quark, three mixing with the top quark and one of electric charge $5/3$. The top quark gained mass purely via the mechanism of partial compositeness while, for the lighter fermions, we resorted to quadratic couplings because of the lack of potential partners.

Much remains to be done before this model can be considered fully satisfactory. The main issue is whether the dynamics of the gauge theory is such that a realistic mass spectrum can be justified. Here we are forced to work at small $\NHC$, so analysis similar to those in~\cite{smallN} could turn out to be useful.
Still, we felt that the IR theory is sufficiently appealing to motivate our study, and thus we presented the most important couplings and discussed the mass spectrum for the top and bottom sector.

The $S$ and $T$ constraints~\cite{EWPT} for this class of models have been discussed in many places and reviewed in~\cite{Contino:2010rs}. These contributions can be made acceptable at the cost of tuning the parameter $\xi$ to be sufficiently small. As for the
$Z \to b \bar b$ decay, we showed that the model belongs to the class of models for which this channel is protected from acquiring large deviations from the SM result. (Top quark compositeness may also constrain these models, see~\cite{Fabbrichesi:2013bca} for an extensive discussion.)

The LHC direct searches during run~1 have probed a large fraction of these models. Limits on the fermionic partners have been set to $m_X \gtrsim 800$~GeV, $m_T \gtrsim  700$~GeV and $m_B \gtrsim  700$~GeV for the charge $5/3$~\cite{Chatrchyan:2013wfa}, charge $2/3$~\cite{Chatrchyan:2013uxa} and charge $-1/3$~\cite{CMS:2013una} respectivelly\footnote{All of these bounds are somewhat dependent on the assumed BR for the decays. We refer to the original literature for a detailed discussion.}.
Searches for doubly charged Higgs bosons appeared in~\cite{ATLAS:2012hi} with the data at 7~TeV, setting a bound of $m_\phi \gtrsim 400$~GeV in various dilepton channels. The search for a generic scalar color octet (called s8 in~\cite{oct}) has excluded a mass range in the region
between 1. and 2.66~TeV, but this limit needs to be analyzed carefully before applying it to the octet appearing in this paper. Lastly, the search for vector resonances~\cite{trho}, also expected to appear at a scale $\Lambda$ in the strongly coupled sector, has set a bound of $m_\rho \gtrsim 1.1$~TeV.
The next LHC run will probe even deeper into these classes of models, exceeding the TeV limit for all composite fermions~\cite{Agashe:2013hma} and ensuring plenty of entertainment for the coming years.

\subsection*{Acknowledgments}

I wish to thank D.~Karateev, R.~Rattazzi, F.~Sannino and M.~Serone for discussion. This research is supported in part by the Swedish Research Council (Vetenskapsr{\aa}det) contract B0508101.


\begin{thebibliography}{99}

%\cite{LHC}
\bibitem{LHC}
  G.~Aad {\it et al.}  [ATLAS Collaboration],
  %``Observation of a new particle in the search for the Standard Model Higgs boson with the ATLAS detector at the LHC,''
  Phys.\ Lett.\ B {\bf 716} (2012) 1
  [arXiv:1207.7214 [hep-ex]].
  %%CITATION = ARXIV:1207.7214;%%
%
  S.~Chatrchyan {\it et al.}  [CMS Collaboration],
  %``Observation of a new boson at a mass of 125 GeV with the CMS experiment at the LHC,''
  Phys.\ Lett.\ B {\bf 716} (2012) 30
  [arXiv:1207.7235 [hep-ex]].
  %%CITATION = ARXIV:1207.7235;%%

  %\cite{BEH}
\bibitem{BEH}
  F.~Englert and R.~Brout,
  %``Broken Symmetry and the Mass of Gauge Vector Mesons,''
  Phys.\ Rev.\ Lett.\  {\bf 13} (1964) 321.
  %%CITATION = PRLTA,13,321;%%
%
  P.~W.~Higgs,
  %``Broken Symmetries and the Masses of Gauge Bosons,''
  Phys.\ Rev.\ Lett.\  {\bf 13} (1964) 508.
  %%CITATION = PRLTA,13,508;%%
%
  G.~S.~Guralnik, C.~R.~Hagen and T.~W.~B.~Kibble,
  %``Global Conservation Laws and Massless Particles,''
  Phys.\ Rev.\ Lett.\  {\bf 13} (1964) 585.
  %%CITATION = PRLTA,13,585;%%

  %\cite{Kaplan:1983fs}
\bibitem{Kaplan:1983fs}
  D.~B.~Kaplan and H.~Georgi,
  %``SU(2) x U(1) Breaking by Vacuum Misalignment,''
  Phys.\ Lett.\ B {\bf 136} (1984) 183.
  %%CITATION = PHLTA,B136,183;%%

%\cite{Kaplan:1991dc}
\bibitem{Kaplan:1991dc}
  D.~B.~Kaplan,
  %``Flavor at SSC energies: A New mechanism for dynamically generated fermion masses,''
  Nucl.\ Phys.\ B {\bf 365} (1991) 259.
  %%CITATION = NUPHA,B365,259;%%

%\cite{CCWZ}
\bibitem{CCWZ}
  S.~R.~Coleman, J.~Wess and B.~Zumino,
  %``Structure of phenomenological Lagrangians. 1.,''
  Phys.\ Rev.\  {\bf 177} (1969) 2239.
  %%CITATION = PHRVA,177,2239;%%
%
  C.~G.~Callan, Jr., S.~R.~Coleman, J.~Wess and B.~Zumino,
  %``Structure of phenomenological Lagrangians. 2.,''
  Phys.\ Rev.\  {\bf 177} (1969) 2247.
  %%CITATION = PHRVA,177,2247;%%

%\cite{manyreviews}
\bibitem{manyreviews}
T.~Gherghetta,
  %``Les Houches lectures on warped models and holography,''
  hep-ph/0601213.
  %%CITATION = HEP-PH/0601213;%%
%
  M.~Schmaltz and D.~Tucker-Smith,
  %``Little Higgs review,''
  Ann.\ Rev.\ Nucl.\ Part.\ Sci.\  {\bf 55} (2005) 229
  [hep-ph/0502182].
  %%CITATION = HEP-PH/0502182;%%
%
 R.~Sundrum,
  %``Tasi 2004 lectures: To the fifth dimension and back,''
  hep-th/0508134.
  %%CITATION = HEP-TH/0508134;%%
%
  M.~Serone,
  %``Holographic Methods and Gauge-Higgs Unification in Flat Extra Dimensions,''
  New J.\ Phys.\  {\bf 12} (2010) 075013
  [arXiv:0909.5619 [hep-ph]].
  %%CITATION = ARXIV:0909.5619;%%
%
  H.~-C.~Cheng,
  %``2009 TASI Lecture -- Introduction to Extra Dimensions,''
  arXiv:1003.1162 [hep-ph].
  %%CITATION = ARXIV:1003.1162;%%
%
C.~Csaki,
  %``TASI lectures on extra dimensions and branes,''
  In *Shifman, M. (ed.) et al.: From fields to strings, vol. 2* 967-1060
  [hep-ph/0404096].
  %%CITATION = HEP-PH/0404096;%%
%
  G.~D.~Kribs,
  %``TASI 2004 lectures on the phenomenology of extra dimensions,''
  hep-ph/0605325.
  %%CITATION = HEP-PH/0605325;%%
%
  R.~Rattazzi,
  %``Cargese lectures on extra-dimensions,''
  *Cargese 2003, Particle physics and cosmology* 461-517
  [hep-ph/0607055].
  %%CITATION = HEP-PH/0607055;%%
%
  C.~Csaki, J.~Hubisz and P.~Meade,
  %``TASI lectures on electroweak symmetry breaking from extra dimensions,''
  hep-ph/0510275.
  %%CITATION = HEP-PH/0510275;%%
%
  B.~Bellazzini, C.~Csáki and J.~Serra,
  %``Composite Higgses,''
  arXiv:1401.2457 [hep-ph].
  %%CITATION = ARXIV:1401.2457;%%

\bibitem{Contino:2010rs}
  R.~Contino,
  %``The Higgs as a Composite Nambu-Goldstone Boson,''
  arXiv:1005.4269 [hep-ph].
  %%CITATION = ARXIV:1005.4269;%%

%\cite{Caracciolo:2012je}
\bibitem{Caracciolo:2012je}
  F.~Caracciolo, A.~Parolini and M.~Serone,
  %``UV Completions of Composite Higgs Models with Partial Compositeness,''
  JHEP {\bf 1302} (2013) 066
  [arXiv:1211.7290 [hep-ph]].
  %%CITATION = ARXIV:1211.7290;%%
%
  D.~Marzocca, A.~Parolini and M.~Serone,
  %``Supersymmetry with a pNGB Higgs and Partial Compositeness,''
  JHEP {\bf 1403} (2014) 099
  [arXiv:1312.5664 [hep-ph]].
  %%CITATION = ARXIV:1312.5664;%%


%\cite{Luty:2004ye}
\bibitem{Luty:2004ye}
  M.~A.~Luty and T.~Okui,
  %``Conformal technicolor,''
  JHEP {\bf 0609} (2006) 070
  [hep-ph/0409274].
  %%CITATION = HEP-PH/0409274;%%

%\cite{smallN}
\bibitem{smallN}
  D.~D.~Dietrich, F.~Sannino and K.~Tuominen,
  %``Light composite Higgs from higher representations versus electroweak precision measurements: Predictions for CERN LHC,''
  Phys.\ Rev.\ D {\bf 72} (2005) 055001
  [hep-ph/0505059].
  %%CITATION = HEP-PH/0505059;%%
%
  A.~Hietanen, R.~Lewis, C.~Pica and F.~Sannino,
  %``Fundamental Composite Higgs Dynamics on the Lattice: SU(2) with Two Flavors,''
  arXiv:1404.2794 [hep-lat].

%\cite{higher}
\bibitem{higher}
  D.~K.~Hong, S.~D.~H.~Hsu and F.~Sannino,
  %``Composite Higgs from higher representations,''
  Phys.\ Lett.\ B {\bf 597} (2004) 89
  [hep-ph/0406200].
  %%CITATION = HEP-PH/0406200;%%
%
  D.~D.~Dietrich and F.~Sannino,
  %``Conformal window of SU(N) gauge theories with fermions in higher dimensional representations,''
  Phys.\ Rev.\ D {\bf 75} (2007) 085018
  [hep-ph/0611341].
  %%CITATION = HEP-PH/0611341;%%

%\cite{confo}
\bibitem{confo}
  R.~Rattazzi, V.~S.~Rychkov, E.~Tonni and A.~Vichi,
  %``Bounding scalar operator dimensions in 4D CFT,''
  JHEP {\bf 0812} (2008) 031
  [arXiv:0807.0004 [hep-th]].
  %%CITATION = ARXIV:0807.0004;%%
%
  V.~S.~Rychkov and A.~Vichi,
  %``Universal Constraints on Conformal Operator Dimensions,''
  Phys.\ Rev.\ D {\bf 80} (2009) 045006
  [arXiv:0905.2211 [hep-th]].
  %%CITATION = ARXIV:0905.2211;%%

%\cite{Barnard:2013zea}
\bibitem{Barnard:2013zea}
  J.~Barnard, T.~Gherghetta and T.~S.~Ray,
  %``UV descriptions of composite Higgs models without elementary scalars,''
  arXiv:1311.6562 [hep-ph].
  %%CITATION = ARXIV:1311.6562;%%

%\cite{Ferretti:2013kya}
\bibitem{Ferretti:2013kya}
  G.~Ferretti and D.~Karateev,
  %``Fermionic UV completions of Composite Higgs models,''
  JHEP {\bf 1403} (2014) 077
  [arXiv:1312.5330 [hep-ph]].
  %%CITATION = ARXIV:1312.5330;%%

%\cite{DeSimone:2012fs}
\bibitem{DeSimone:2012fs}
  A.~De Simone, O.~Matsedonskyi, R.~Rattazzi and A.~Wulzer,
  %``A First Top Partner Hunter's Guide,''
  JHEP {\bf 1304} (2013) 004
  [arXiv:1211.5663 [hep-ph]].
  %%CITATION = ARXIV:1211.5663;%%

%\cite{Carena:2014ria}
\bibitem{Carena:2014ria}
  M.~Carena, L.~Da Rold and E.~Ponton,
  %``Minimal Composite Higgs Models at the LHC,''
  arXiv:1402.2987 [hep-ph].
  %%CITATION = ARXIV:1402.2987;%%

%\cite{Agashe:2006at}
\bibitem{Agashe:2006at}
  K.~Agashe, R.~Contino, L.~Da Rold and A.~Pomarol,
  %``A Custodial symmetry for Zb anti-b,''
  Phys.\ Lett.\ B {\bf 641} (2006) 62
  [hep-ph/0605341].
  %%CITATION = HEP-PH/0605341;%%

%\cite{oai:arXiv.org:0910.1789}
\bibitem{oai:arXiv.org:0910.1789}
  B.~Gripaios,
  %``Composite Leptoquarks at the LHC,''
  JHEP {\bf 1002} (2010) 045
  [arXiv:0910.1789 [hep-ph]].
  %%CITATION = ARXIV:0910.1789;%%

%\cite{'tHooft:1980xb}
\bibitem{'tHooft:1980xb}
  G.~'t Hooft,
  %``Recent Developments in Gauge Theories. Proceedings, Nato Advanced Study Institute, Cargese, France, August 26 - September 8, 1979,''
  NATO Adv.\ Study Inst.\ Ser.\ B Phys.\  {\bf 59} (1980) pp.1.
  %%CITATION = NASBD,59,pp.1;%%

%\cite{Preskill:1981sr}
\bibitem{Preskill:1981sr}
  J.~Preskill and S.~Weinberg,
  %``'decoupling' Constraints On Massless Composite Particles,''
  Phys.\ Rev.\ D {\bf 24} (1981) 1059.
  %%CITATION = PHRVA,D24,1059;%%

%\cite{VW}
\bibitem{VW}
  C.~Vafa and E.~Witten,
  %``Restrictions on Symmetry Breaking in Vector-Like Gauge Theories,''
  Nucl.\ Phys.\ B {\bf 234} (1984) 173.
  %%CITATION = NUPHA,B234,173;%%

%\cite{NJL}
\bibitem{NJL}
  Y.~Nambu and G.~Jona-Lasinio,
  %``Dynamical Model of Elementary Particles Based on an Analogy with Superconductivity. 1.,''
  Phys.\ Rev.\  {\bf 122} (1961) 345.
  %%CITATION = PHRVA,122,345;%%
%
  Y.~Nambu and G.~Jona-Lasinio,
  %``Dynamical Model Of Elementary Particles Based On An Analogy With Superconductivity. Ii,''
  Phys.\ Rev.\  {\bf 124} (1961) 246.
  %%CITATION = PHRVA,124,246;%%

%\cite{Raby:1979my}
\bibitem{Raby:1979my}
  S.~Raby, S.~Dimopoulos and L.~Susskind,
  %``Tumbling Gauge Theories,''
  Nucl.\ Phys.\ B {\bf 169} (1980) 373.
  %%CITATION = NUPHA,B169,373;%%

%\cite{Georgi:1981mh}
\bibitem{Georgi:1981mh}
  H.~Georgi, L.~J.~Hall and M.~B.~Wise,
  %``Remarks on Mass Hierarchies From Tumbling Gauge Theories,''
  Phys.\ Lett.\ B {\bf 102} (1981) 315
   [Erratum-ibid.\ B {\bf 104} (1981) 499].
  %%CITATION = PHLTA,B102,315;%%

%\cite{Eichten:1981mu}
\bibitem{Eichten:1981mu}
  E.~Eichten and F.~Feinberg,
  %``Comment on Tumbling Gauge Theories,''
  Phys.\ Lett.\ B {\bf 110} (1982) 232.
  %%CITATION = PHLTA,B110,232;%%

%\cite{Dugan:1984hq}
\bibitem{Dugan:1984hq}
  M.~J.~Dugan, H.~Georgi and D.~B.~Kaplan,
  %``Anatomy of a Composite Higgs Model,''
  Nucl.\ Phys.\ B {\bf 254} (1985) 299.
  %%CITATION = NUPHA,B254,299;%%

%\cite{Georgi:1984af}
\bibitem{Georgi:1984af}
  H.~Georgi and D.~B.~Kaplan,
  %``Composite Higgs and Custodial SU(2),''
  Phys.\ Lett.\ B {\bf 145} (1984) 216.
  %%CITATION = PHLTA,B145,216;%%

%\cite{Georgi:1985nv}
\bibitem{Georgi:1985nv}
  H.~Georgi and M.~Machacek,
  %``Doubly Charged Higgs Bosons,''
  Nucl.\ Phys.\ B {\bf 262} (1985) 463.
  %%CITATION = NUPHA,B262,463;%%

%\cite{Agashe:2004rs}
\bibitem{Agashe:2004rs}
  K.~Agashe, R.~Contino and A.~Pomarol,
  %``The Minimal composite Higgs model,''
  Nucl.\ Phys.\ B {\bf 719} (2005) 165
  [hep-ph/0412089].
  %%CITATION = HEP-PH/0412089;%%

%\cite{Witten:1983ut}
\bibitem{Witten:1983ut}
  E.~Witten,
  %``Some Inequalities Among Hadron Masses,''
  Phys.\ Rev.\ Lett.\  {\bf 51} (1983) 2351.
  %%CITATION = PRLTA,51,2351;%%

%\cite{Coleman:1973jx}
\bibitem{Coleman:1973jx}
  S.~R.~Coleman and E.~J.~Weinberg,
  %``Radiative Corrections as the Origin of Spontaneous Symmetry Breaking,''
  Phys.\ Rev.\ D {\bf 7} (1973) 1888.
  %%CITATION = PHRVA,D7,1888;%%

%\cite{Vecchi:2013bja}
\bibitem{Vecchi:2013bja}
  L.~Vecchi,
  %``The Natural Composite Higgs,''
  arXiv:1304.4579 [hep-ph].
  %%CITATION = ARXIV:1304.4579;%%

%\cite{Katz:2005au}
\bibitem{Katz:2005au}
  E.~Katz, A.~E.~Nelson and D.~G.~E.~Walker,
  %``The Intermediate Higgs,''
  JHEP {\bf 0508} (2005) 074
  [hep-ph/0504252].
  %%CITATION = HEP-PH/0504252;%%

%\cite{Gripaios:2009pe}
\bibitem{Gripaios:2009pe}
  B.~Gripaios, A.~Pomarol, F.~Riva and J.~Serra,
  %``Beyond the Minimal Composite Higgs Model,''
  JHEP {\bf 0904} (2009) 070
  [arXiv:0902.1483 [hep-ph]].
  %%CITATION = ARXIV:0902.1483;%%

%\cite{Galloway:2010bp}
\bibitem{Galloway:2010bp}
  J.~Galloway, J.~A.~Evans, M.~A.~Luty and R.~A.~Tacchi,
  %``Minimal Conformal Technicolor and Precision Electroweak Tests,''
  JHEP {\bf 1010} (2010) 086
  [arXiv:1001.1361 [hep-ph]].
  %%CITATION = ARXIV:1001.1361;%%

%\cite{EWPT}
\bibitem{EWPT}
  G.~Altarelli and R.~Barbieri,
  %``Vacuum polarization effects of new physics on electroweak processes,''
  Phys.\ Lett.\ B {\bf 253} (1991) 161.
  %%CITATION = PHLTA,B253,161;%%
%
  M.~E.~Peskin and T.~Takeuchi,
  %``A New constraint on a strongly interacting Higgs sector,''
  Phys.\ Rev.\ Lett.\  {\bf 65} (1990) 964.
  %%CITATION = PRLTA,65,964;%%
%
  M.~E.~Peskin and T.~Takeuchi,
  %``Estimation of oblique electroweak corrections,''
  Phys.\ Rev.\ D {\bf 46} (1992) 381.
  %%CITATION = PHRVA,D46,381;%%

%\cite{Fabbrichesi:2013bca}
\bibitem{Fabbrichesi:2013bca}
  M.~Fabbrichesi, M.~Pinamonti and A.~Tonero,
  %``Stringent limits on top-quark compositeness from top anti-top production at the Tevatron and the LHC,''
  Phys.\ Rev.\ D {\bf 89} (2014) 074028
  [arXiv:1307.5750 [hep-ph]].
  %%CITATION = ARXIV:1307.5750;%%

%\cite{Chatrchyan:2013wfa}
\bibitem{Chatrchyan:2013wfa}
  S.~Chatrchyan {\it et al.}  [CMS Collaboration],
  %``Search for top-quark partners with charge 5/3 in the same-sign dilepton final state,''
  arXiv:1312.2391 [hep-ex].
  %%CITATION = ARXIV:1312.2391;%%

%\cite{Chatrchyan:2013uxa}
\bibitem{Chatrchyan:2013uxa}
  S.~Chatrchyan {\it et al.}  [CMS Collaboration],
  %``Inclusive search for a vector-like T quark with charge $\frac{2}{3}$ in pp collisions at $\sqrt{s}$ = 8 TeV,''
  Phys.\ Lett.\ B {\bf 729} (2014) 149
  [arXiv:1311.7667 [hep-ex]].
  %%CITATION = ARXIV:1311.7667;%%

%\cite{CMS:2013una}
\bibitem{CMS:2013una}
  CMS Collaboration [CMS Collaboration],
  %``Search for Vector-Like b' Pair Production with Multilepton Final States in pp collisions at sqrt(s) = 8 TeV,''
  CMS-PAS-B2G-13-003.
  %%CITATION = CMS-PAS-B2G-13-003;%%

%\cite{ATLAS:2012hi}
\bibitem{ATLAS:2012hi}
  G.~Aad {\it et al.}  [ATLAS Collaboration],
  %``Search for doubly-charged Higgs bosons in like-sign dilepton final states at $\sqrt{s}=7$ TeV with the ATLAS detector,''
  Eur.\ Phys.\ J.\ C {\bf 72} (2012) 2244
  [arXiv:1210.5070 [hep-ex]].
  %%CITATION = ARXIV:1210.5070;%%

%\cite{oct}
\bibitem{oct}
  T.~Han, I.~Lewis and Z.~Liu,
  %``Colored Resonant Signals at the LHC: Largest Rate and Simplest Topology,''
  JHEP {\bf 1012} (2010) 085
  [arXiv:1010.4309 [hep-ph]].
  %%CITATION = ARXIV:1010.4309;%%
%
  CMS Collaboration [CMS Collaboration],
  %``Search for Narrow Resonances using the Dijet Mass Spectrum in pp Collisions at sqrt s of 8 TeV,''
  CMS-PAS-EXO-12-016.
  %%CITATION = CMS-PAS-EXO-12-016;%%

%\cite{trho}
\bibitem{trho}
  CMS Collaboration [CMS Collaboration],
  %``Search for W'/technirho in WZ using leptonic final states,''
  CMS-PAS-EXO-12-025.
  %%CITATION = CMS-PAS-EXO-12-025;%%
%
[ATLAS Collaboration],
  %``Search for resonant WZ → 3ℓ ν production in 1 √s = 8 TeV pp collisions with 13 fb−1 at ATLAS,''
  ATLAS-CONF-2013-015.
  %%CITATION = ATLAS-CONF-2013-015;%%

%\cite{Agashe:2013hma}
\bibitem{Agashe:2013hma}
  K.~Agashe {\it et al.}  [Top Quark Working Group Collaboration],
  %``Snowmass 2013 Top quark working group report,''
  arXiv:1311.2028 [hep-ph].
  %%CITATION = ARXIV:1311.2028;%%

\end{thebibliography}
\end{document}